\documentclass[a4paper,12pt]{article}

\usepackage{amsmath,amsthm,amsbsy,amssymb,amsfonts,graphics,ascmac,mathrsfs}
\usepackage[dvipdfmx]{graphicx,color}

\makeatletter
\catcode`\@=11
\@addtoreset{equation}{section}

\makeatother

\baselineskip=\normalbaselineskip

\allowdisplaybreaks[3]
\renewcommand{\thefootnote}{\arabic{footnote}}
\newcommand{\gs}{g_s}
\newcommand{\ls}{l_s}
\newcommand{\Exp}[1]{\operatorname{e}^{\, #1}}
\newcommand{\abs}[1]{{\lvert #1 \rvert}}
\newcommand{\rmd}{{\mathrm{d}}}
\newcommand{\ii}{{\hspace{0.7pt}\mathrm{i}\hspace{0.7pt}}}
\newcommand{\Tr}{\operatorname{Tr}}
\newcommand{\nn}{\nonumber}
\def\bpm{\begin{pmatrix}}
\def\epm{\end{pmatrix}}
\newcommand\smatrix[1]{\begin{footnotesize}\begin{pmatrix}#1\end{pmatrix}\end{footnotesize}}

\def\cA{\mathcal A}
\def\cC{\mathcal C}
\def\cF{\mathcal F}
\def\cG{\mathcal G}\def\cH{\mathcal H}

\newcommand{\lR}{{\mathbb R}}
\newcommand{\lZ}{{\mathbb Z}}
\newcommand{\SLa}{\mathsf{a}}
\newcommand{\SLb}{\mathsf{b}}
\newcommand{\SLc}{\mathsf{c}}
\newcommand{\SLd}{\mathsf{d}}
\newcommand{\axion}{\chi}
\newcommand{\KKM}{{\rm KKM}}
\newcommand{\NS}{{\rm NS}}
\newcommand{\FF}{{\rm F}}
\newcommand{\DD}{{\rm D}}
\newcommand{\PP}{{\rm P}}
\newcommand{\LL}{L}

\begin{document}

\begin{titlepage}
\renewcommand{\thefootnote}{\fnsymbol{footnote}}

\begin{flushright}
\parbox{3.5cm}
{SNUTP14-012}
\end{flushright}

\vspace*{1.0cm}

\begin{center}
{\Large \bf Defect branes as Alice strings}%
\end{center}
\vspace{1.0cm}

\centerline{
{Takashi Okada}%
\footnote{E-mail address: 
takashi.okada@riken.jp} 
and
{Yuho Sakatani}%
\footnote{E-mail address: 
yuho@cc.kyoto-su.ac.jp}%
}

\vspace{0.2cm}

\begin{center}
${}^\ast${\it Theoretical Biology Laboratory, RIKEN, \\
Wako 351-0198, JAPAN}

${}^\dagger${\it Department of Physics and Astronomy, \\
Seoul National University, Seoul 151-747, KOREA\\}

\end{center}
\vspace*{1cm}
\begin{abstract}

There exist various defect-brane backgrounds in supergravity theories 
which arise as the low energy limit of string theories. 
These backgrounds typically have non-trivial monodromies, 
and if we move a charged probe around the center of a defect, 
its charge will be changed by the action of the monodromy. 
During the process, the charge conservation law seems to be violated. 
In this paper, to resolve this puzzle, 
we examine a dynamics of the charge changing process 
and show that the missing charge of the probe is transferred to the background. 
We then explicitly construct the resultant background 
after the charge transfer process by utilizing dualities. 
This background has the same monodromy as the original defect brane, 
but has an additional charge which does not have any localized source. 
In the literature, such a charge without localized source 
is known to appear in the presence of Alice strings. 
We argue that defect branes can in fact be regarded as 
a realization of Alice strings in string theory 
and examine the charge transfer process from that perspective. 

\end{abstract}
\thispagestyle{empty}
\end{titlepage}

\setcounter{footnote}{0}

\section{Introduction}
\label{sec:introduction}

String theory has various defect branes (or codimension-two branes), 
including the well-known D7-brane and various exotic branes. 
In the supergravity description, 
some of the corresponding backgrounds are called 
non-geometric backgrounds or U-folds \cite{Hull:2004in}, 
since the transition functions between coordinate patches 
are given by U-duality transformations. 
In particular, the authors in \cite{deBoer:2010ud} pointed out that 
the background of $5_2^2$-branes \cite{LozanoTellechea:2000mc} has a T-duality monodromy 
and it is a concrete example of non-geometric backgrounds in string theory. 

The non-trivial monodromies of defect branes, which is familiar for a D7-brane, 
generally raise a perplexing problem related to the charge conservation law. 
Let us consider a charged probe brane in a defect-brane background. 
If we move the probe around the center of the defect, 
its charge will change due to the action of the monodromy. 
Where does the original charge of the probe go 
and how is the charge conservation law kept intact?
In \cite{deBoer:2012ma}, it was proposed that the charge is indeed conserved 
if we measure the charge by using the Page charge 
\cite{Page:1984qv,Marolf:2000cb,deBoer:2012ma}, 
which is one of the possible definitions of charge. 
However, in defect-brane backgrounds, the definition 
of the (Page) charge depends on the choice of a cycle 
for a flux integral (see \cite{deBoer:2012ma} and section \ref{sec:defect-alice}). 
Thus, it is desirable to define the charge in a consistent manner 
to analyze the (apparent) charge changing phenomena. 

In the literature, the similar issue was discussed 
in a certain class of $(1+3)$-dimensional gauge theories 
which admit vortex solutions with non-trivial monodromies 
\cite{Schwarz:1982ec,Alford:1990mk,Preskill:1990bm}. 
One of the most famous and studied example is a vortex 
called an Alice string \cite{Schwarz:1982ec}, whose monodromy is given by a charge conjugation.%%%
\footnote{The physics of Alice string and the variants has been studied in various fields from cosmological physics \cite{Blinnikov:1982eh,Brekke:1991ap,BenMenahem:1992fe,Kobayashi:2010na} to condensed matter physics \cite{Stephen:1974ur,Wright:1989zz}.} 
That is, after a particle with charge $q$ goes around an Alice string, 
the sign of its charge flips. 
In the presence of Alice strings, ``a charge with no localized source,'' 
called a Cheshire charge \cite{Alford:1990mk,Preskill:1990bm}, 
plays an important role in the discussion of the charge conservation law. 
In this paper, we will examine an analogy between Alice strings 
and defect branes in string theory 
(see \cite{Harvey:2007ab} for an earlier study 
on a realization of Alice string in string theory). 

In order to discuss the issue of the charge conservation law 
in defect-brane backgrounds more concretely, 
let us consider a Kaluza-Klein (KK) vortex (or smeared KK monopole) background
\cite{Meessen:1998qm,Onemli:2000kb,LozanoTellechea:2000mc} as an example. 
A key feature of the KK-vortex background different from other defect-brane backgrounds 
is that the monodromy of the KK-vortex background is just a coordinate transformation, 
which enables us to understand the charge changing phenomena geometrically. 
The KK-vortex background is given by
\begin{align}
 \rmd s^2&= -\rmd t^2 + H(r)\, \bigl(\rmd r^2+r^2\,\rmd\theta^2+\rmd x_3^2\bigr) 
            + H^{-1}(r)\,\bigl[\rmd x^4 - \sigma\,(\theta/2\pi) \, \rmd x^3\bigr]^2 
            + \rmd x^2_{56789} \,, 
\\
 \Exp{2\hat{\phi}} &= 1\,,\qquad \hat{B}^{(2)} = 0 \,, \quad 
 H(r)\equiv (\sigma/2\pi)\,\log(r_{\rm c}/r) \,.
\label{eq:smeared-KKM}
\end{align}
Here, we defined $\sigma \equiv R_4/R_3$ where $R_i$ ($i=3,\dotsc,9$) 
is the compactification radius in the $x^i$-direction, 
and $r_{\rm c}$ is a cutoff radius of the geometry; 
the geometry gives a good description only for $r\ll r_{\rm c}$.%%%%
\footnote{The geometry also has a singularity near the center, which can be resolved in string theory (see e.g.~\cite{Onemli:2000kb}).}
If we gather the $x^3$-$x^4$ components of the metric and $B$-field 
into the generalized metric of a $4\times 4$ matrix
\begin{align}
 \cH^{-1} 
 \equiv \bpm \hat{G}^{-1} & -\hat{G}^{-1}\,\hat{B} \cr 
    \hat{B}\,\hat{G}^{-1} & \hat{G}-\hat{B}\,\hat{G}^{-1}\,\hat{B} \epm \,, 
\label{eq:34generalized-metric}
\end{align}
the monodromy around the center, $r=0$, is given by the matrix $\Omega_{\KKM}$:
\begin{align}
 &\cH^{-1}(\theta=2\pi) 
 = \Omega_{\KKM}^{\rm T} \,\cH^{-1}(\theta=0) \,\Omega_{\KKM}\,,
\\
 &\Omega_{\KKM} \equiv 
 \bpm 
 {\boldsymbol\omega}^{\rm T} & \mathbf{0} \cr 
 \mathbf{0}  & {\boldsymbol\omega}^{-1} \cr 
 \epm \,,\quad 
 {\boldsymbol\omega}
 \equiv \bpm 1& 0 \cr 
 \sigma & \,1\, \cr 
 \epm\,.
\end{align}
The monodromy matrix $\Omega_{\KKM}$ characterizes 
the existence of the KK vortex at the center. 

\begin{figure}[tb]
 \centering
 \includegraphics[width=11cm]{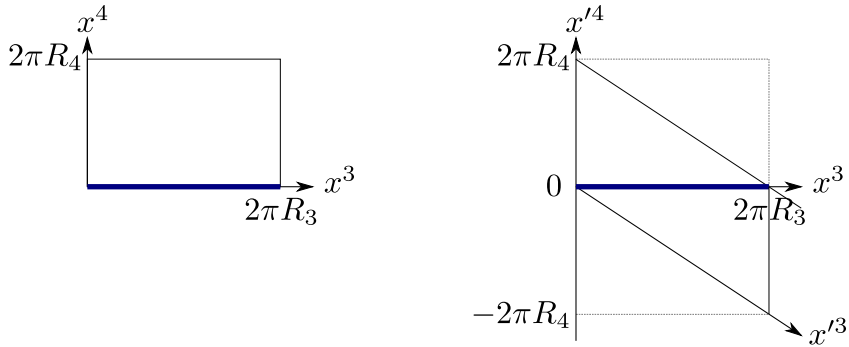}
 \caption{Coordinate system in the 3-4 torus at $\theta=0$ (left) and $\theta=2\pi$ (right). 
 The thick line represents the portrait of the string.} 
\label{fig:winding}
\end{figure}

Due to the existence of the non-trivial monodromy, 
if we put a probe string with $\FF1(3)$ charge%%%
\footnote{See Appendix \ref{app:conventions} for the notation of various brane charges.}
and move it once around the center counterclockwise, 
its charge will change as
\begin{align}
 \smatrix{(1/R_3)\times\#\PP(3)\cr (1/R_4)\times\#\PP(4)\cr (R_3/\ls^2)\times\#\FF1(3) \cr (R_4/\ls^2)\times\#\FF1(4)}
 = \smatrix{0\cr 0\cr (R_3/\ls^2)\times 1 \cr 0} 
 \ \to \ 
   \Omega_{\KKM}^{-1}\,\smatrix{0\cr 0\cr (R_3/\ls^2)\times 1 \cr 0} 
 = \smatrix{0\cr 0\cr (R_3/\ls^2)\times 1 \cr (R_4/\ls^2)\times 1} \,.
\label{eq:KKM-monodromy-charge}
\end{align}
The change in the winding charge can be understood geometrically; 
the monodromy $\Omega_{\KKM}$ of the KK vortex 
corresponds to the diffeomorphism on a 3-4 torus:
\begin{align}
 x^{\prime 3}=x^3\,,\qquad x^{\prime 4}=x^4 + \sigma\,x^3\,.
\label{eq:coordinate-change}
\end{align}
As described in Figure \ref{fig:winding} (left), 
the probe is initially extending along the $x^3$-direction. 
After it goes around the center of the defect, in the above primed coordinates, 
the probe string extends from $(x^{\prime 3},x^{\prime 4})=(0,\,0)$ to $(2\pi R_3,\,2\pi R_4)$\,.
That is, if we count the winding number using the primed coordinates, 
the resulting charge is $\FF1(3)+\FF1(4)$, which agrees with \eqref{eq:KKM-monodromy-charge}. 
The reason why we use the primed coordinates 
for the purpose of measuring the winding charge will be discussed in section \ref{sec:defect-alice}. 

If the total charge is to be conserved, we expect that the background after the probe rotated should 
carry a flux that compensates the charge change of the probe. 
In this paper, we examine in detail the dynamical process 
in which the winding charge of a probe string is transferred to the background, 
and discuss on the charge conservation during this process. 
Then, we explicitly construct a deformed defect-brane background with an additional charge 
by utilizing the four-dimensional electric-magnetic duality. 
We show that the additional charge is exactly the same as that left behind by the probe brane 
and thus the background can be regarded as the resultant background after the probe goes around the defect.%%%
\footnote{See \cite{Giusto:2010gv} for a work with a similar motivation; the authors considered the unwinding process in a fuzzball geometry and constructed a fuzzball geometry with a non-trivial field strength.}

We also examine the charge transfer process in the $5^2_2$-brane background. 
Compared with the KK-vortex, the case is rather difficult to understand geometrically 
since the monodromy is not the usual coordinate transformation but a T-duality transformation. 
 We will circumvent this obstacle by introducing the double field theory 
\cite{Hull:2009mi,Hull:2009zb,Hohm:2010jy,Hohm:2010pp,Hohm:2011dv,Hull:2006va}.

This paper is organized as follows. 
In section \ref{sec:charge-transfer}, we consider a probe string rotating around a KK vortex 
and examine the detailed dynamics of the charge transfer process explained above, 
based on the analysis of \cite{Gregory:1997te}. 
In section \ref{sec:alice-defect}, we review the notions of Alice strings and Cheshire charges 
and discuss their relevance to defect branes in string theory. 
In section \ref{sec:KKM-dyon}, we review the construction of the KK-dyon solution following \cite{Roy:1998ev},
which uses the electric-magnetic duality in four-dimensional theory. 
We then find that this background has F1 charges which do not have localized source, 
just like Cheshire charges. 
Moreover, using a smearing procedure, we construct a dyonic KK-vortex solution and discuss its relevance 
to the charge transfer process considered in section \ref{sec:charge-transfer}. 
In section \ref{sec:522-background}, we first construct the $5^2_2$ background with F1 charges 
(with no localized source) by using a duality transformation and show that it is a $T$-fold, 
whose monodromy is the same as that of the pure $5_2^2$ background. 
Then, we consider the charge transfer process in the $5^2_2$ background, 
and argue that the change of charge can be geometrically understood 
if we describe the $5^2_2$ background as a doubled geometry. 
In section \ref{sec:various-alice}, using successive $U$-dualities, 
we obtain various defect-brane backgrounds with Cheshire charges. 
We then find another construction of such backgrounds 
without the use of the electric-magnetic duality in four dimensions. 
Section \ref{sec:conclusion} is devoted to discussions and conclusion. 

\section{Monodromy and charge transfer process}
\label{sec:charge-transfer}

As we discussed in the introduction, once a probe brane moves around a defect brane, 
it is natural to expect that some charges are transferred from the probe brane to the background. 
In this section, we examine the charge transfer process following the analysis of \cite{Gregory:1997te}. 

Concretely, we consider a probe string with a winding charge, $\FF1(4)$, 
rotating around the center of the KK-vortex background \eqref{eq:smeared-KKM}. 
In the following, we dimensionally reduce the $x^4,\dotsc,x^9$-directions, 
and thus the probe string here is smeared along these directions (i.e.~it has codimension two). 
For generality, we make the following ansatz for the background fields:%%%
\footnote{See Appendix \ref{app:ele-mag} for the detailed definitions of four-dimensional fields.}
\begin{align}
 \rmd s^2&= \hat{G}_{MN}\,\rmd x^M\,\rmd x^N 
 =g_{\mu\nu}\,\rmd x^\mu\,\rmd x^\nu 
           + G_{44}\, \bigl(\rmd x^4+A^4\bigr)^2 + \rmd x^2_{5\cdots9}\,,
\\
 \hat{B}^{(2)} &= \frac{1}{2}\,\hat{B}^{(2)}_{MN}\,\rmd x^M\wedge\rmd x^N
 =\frac{1}{2}\,\bigl(B_{\mu\nu}+A^4_\mu\,A_{4\nu}\bigr)\,\rmd x^\mu\wedge\rmd x^\nu
               + A_4 \wedge \bigl(\rmd x^4+A^4\bigr)
\nn\\
 \bigl[A^4 &= A^4_\mu\,\rmd x^\mu \,,\quad 
       A_4 = A_{4\mu}\,\rmd x^\mu\,,\quad 
       \cA^I \equiv \bigl(A^4\,,\ A_4\bigr)^{\rm T}\,,\quad
       \mu,\,\nu=0,\dotsc,3\,\bigr]\,.
\end{align}
By considering a compactification to four dimensions, 
we obtain the following action:
\begin{align}
 S &= S_{\rm bulk} + S_{\rm probe} \,,
\\
 S_{\rm bulk}&\equiv \frac{1}{2\kappa_4^2}\,\int \rmd^4x\,\sqrt{-g}\Exp{-2\phi}
 \Bigl[R -\frac{1}{12}\,H_{\mu\nu\rho}\,H^{\mu\nu\rho}
       -\frac{1}{4}(M^{-1})_{IJ}\,\cF^I_{\mu\nu}\,\cF^{J\mu\nu}+\cdots \Bigr] \,,
\\
 S_{\rm probe}&\equiv -\frac{1}{4\pi\ls^2}\,\int\rmd^2\sigma\,\rmd^4 x\,\delta^4(x-X(\boldsymbol{\sigma}))\, 
 \bigl(\eta^{ab}\,\hat{G}_{MN} + \epsilon^{ab}\,\hat{B}^{(2)}_{MN}\bigr) \,\partial_a X^M\, \partial_b X^N
\nn\\
 &= -\frac{1}{4\pi\ls^2}\,\int\rmd^2 \sigma \, \rmd^4 x\,\delta^4(x-X(\boldsymbol{\sigma}))
\nn\\
 &\qquad\qquad \times\Bigl[ \eta^{ab}\,\bigl(h_{ab} + G_{44}\,V_a\,V_b \bigr)
  + \epsilon^{ab}\,\bigl(b_{ab} + A^4_a\,A_{4b} +2 A_{4a}\,V_b \bigr)\Bigr] \,,
\end{align}
where $(\boldsymbol{\sigma})=(\sigma^0,\,\sigma^1)$ ($0\leq\sigma^1\leq1$) are worldsheet coordinates and we defined
\begin{align}
 \cF^I&\equiv \rmd \cA^I\,,\quad 
 M^{-1} \equiv \bpm G_{44} & 0 \cr 0 & G_{44}^{-1}\epm \,,\quad 
 H \equiv \rmd B -\frac{1}{2}\,\bigl(\cA^4\wedge \cF_4+\cA_4\wedge \cF^4 \bigr) \,,
\\
 h_{ab}&\equiv g_{\mu\nu}\,\partial_a X^\mu\, \partial_b X^\nu \,,\quad 
 V_a \equiv \partial_a X^4+ A^4_\mu\,\partial_a X^\mu\,,\quad 
 b_{ab}\equiv B_{\mu\nu}\,\partial_a X^\mu\, \partial_b X^\nu \,,
\end{align}
and $\epsilon^{10}=-\epsilon^{01}=1$, 
and assumed $X^\alpha(\boldsymbol{\sigma})=\text{const.}$%%%
\footnote{Note that the embedding functions describe a trajectory of the probe string before smearing.} 
($\alpha=5,\dotsc,9$). 

The equations of motion for $B_{\mu\nu}$ and $\cA^I$ are then given by
\begin{align}
 \partial_\mu\bigl(\sqrt{-g}\,\Exp{-2\phi}H^{\mu\nu\rho}\bigr)
 &= \frac{\kappa_4^2}{\pi\ls^2}\,\int\rmd^2 \sigma \,\delta^4(x-X(\boldsymbol{\sigma}))\,
    \epsilon^{ab}\,\partial_a X^\nu\,\partial_b X^\rho\,,
\\
 \partial_\mu\bigl[\sqrt{-g}\,\Exp{-2\phi}\,(M^{-1})_{IJ}\,\cF^{J\mu\nu}\bigr] 
 &= -\frac{\sqrt{-g}}{2}\,\Exp{-2\phi} H^{\nu\mu\rho} \,\LL_{IJ}\,\cF^J_{\mu\rho}
\nn\\
 &\quad +\frac{\kappa_4^2}{\pi\ls^2}\,\int\rmd^2\sigma\,\delta^4(x-X(\boldsymbol{\sigma}))\,
         T_I^{ab}\,V_a\, \partial_b X^\nu 
\end{align}
with $(T_I^{ab}) \equiv \bigl(G_{44}\,\eta^{ab}\,, \ -\epsilon^{ab}\bigr)^{\rm T}$ and 
$(\LL_{IJ})\equiv \bigl(\begin{smallmatrix} 0&1 \cr 1&0 \end{smallmatrix}\bigr)$\,.

We define a physical electric charge $Q_I$ \cite{Schwarz:1993mg} by
\begin{align}
 Q_I &\equiv \int_{V} *_{\rm 4E} j_I 
 = \frac{1}{2\kappa_4^2}\,\int_{\partial D\times I_3} \Exp{-2\phi}\,(M^{-1})_{IJ}\,*_{\rm 4E}\cF^J 
\nn\\
 \biggl[\, *_{\rm 4E} j_I
  &\equiv \frac{1}{2\kappa_4^2}\,\rmd \bigl[\Exp{-2\phi}\,(M^{-1})_{IJ}\,*_{\rm 4E}\cF^J\bigr]\,,\quad 
 j_I^\nu \equiv \frac{1}{2\kappa_4^2}\,\nabla_\mu \bigl[\Exp{-2\phi}\,(M^{-1})_{IJ}\,\cF^{J\mu\nu}\bigr]\,\biggr] \,,
\end{align}
where $*_{\rm 4E}$ is the Hodge star operator associated with the four-dimensional Einstein frame 
and $V$ is a solid cylinder $D\times I_3$ 
($D$: a disk with the radius $r_\infty$ in the transverse two-dimensional space, 
$I_3$: an interval $[0,2\pi R_3]$ in the $x^3$-direction). 
We also define the duality covariant charge vector by
\begin{align}
 \bpm p^I \cr q_{I} \epm
   \equiv  \bpm -\frac{1}{2\kappa_4^2}\,\int_{\partial D\times I_3} \cF^I \cr 
                -\frac{1}{2\kappa_{4}^2}\,\int_{\partial D\times I_3} \cG_I \epm \,,
\quad 
 \cG_I \equiv - \Exp{-2\phi}\,(M^{-1})_{IJ}\, *_{\rm 4E} \cF^{J} - \axion\,\LL_{IJ} \, \cF^{J}\,,
\end{align}
where $\axion$ is the axion field defined in \eqref{eq:H3-axion}. 
Then, each component of the charge vector $p^4$,\ $p_4$,\ $q^4$, and $q_4$ 
corresponds to KKM(56789,4), NS5(56789), P(4), and F1(4) charge, respectively. 
The units of these charges are given by
\begin{align}
 q_{\KKM} \equiv \frac{2\pi R_4}{2\kappa_4^2} \,,\quad 
 q_{\NS5} \equiv \frac{(2\pi\ls)^2}{2\kappa_4^2\,(2\pi R_4)} \,,
 \quad 
 q_{\PP} \equiv \frac{1}{R_4} \,, \quad
 q_{\FF1}\equiv \frac{2\pi R_4}{2\pi\ls^2}
\end{align}
with $V_{i_1\cdots i_n}\equiv (2\pi R_{i_1})\cdots (2\pi R_{i_n})$\,. 
Note that the difference between the physical and the duality covariant 
electric charge is in the term proportional to the axion. 

With the above definitions, 
the equations of motion for gauge fields can be written as
\begin{align}
  j_I^\mu(x) = -\frac{1}{4\kappa_4^2}\,\Exp{-2\phi} H^{\mu\nu\rho} \,\LL_{IJ}\,\cF^J_{\nu\rho}
 +\frac{1}{2\pi\ls^2}\,\int\rmd^2\sigma\,
  \frac{\delta^4(x-X(\boldsymbol{\sigma}))}{\sqrt{-g}}\,T_I^{ab}\,V_a\, \partial_b X^\mu \,,
\end{align}
or
\begin{align}
  \mathbf{j}_I^\mu(x) &\equiv -\frac{1}{2\kappa_4^2}\,\bigl(*_{\rm 4E}\rmd\cG_I\bigr)^\mu
  = \frac{1}{2\kappa_4^2}\,
    \nabla_\nu \bigl[\Exp{-2\phi}\,(M^{-1})_{IJ}\,\cF^{J\mu\nu}-\axion\,\LL_{IJ}\,\tilde{\cF}^{\mu\nu}\bigr]
\nn\\
 &= \frac{1}{2\pi\ls^2}\,\int\rmd^2\sigma\,
    \frac{\delta^4(x-X(\boldsymbol{\sigma}))}{\sqrt{-g}}\,T_I^{ab}\,V_a\, \partial_b X^\mu \,.
\end{align}
The current $\mathbf{j}_I^\mu(x)$ counts only the brane source charge 
while the physical current $j_I^\mu(x)$ additionally includes the charges dissolved into the flux. 

\begin{figure}[h]
 \centering
\includegraphics[width=7cm]{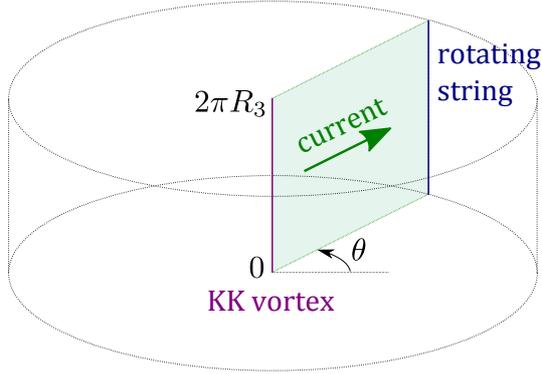}
\caption{Outflow of the $\FF1(4)$ charge from the KK vortex to the rotating probe string. 
The vertical direction is $x^3$, which has a period $2\pi R_3$.} 
\label{fig:rotating-string}
\end{figure}

Now, we consider the following trajectory of the probe string:
\begin{align}
 X^t=\sigma^0\,,\quad X^r = r_0\,,\quad X^\theta = 2\pi \sigma^0\,,\quad 
 X^3=2\pi R_3\,\sigma^1\,,\quad X^4=x^4=\text{const.} 
\label{eq:trajectry} 
\end{align}
By assuming that the time-derivative term 
$\partial_t \bigl(\sqrt{-g}\,\Exp{-2\phi}H^{t\theta 3}\bigr)$ can be neglected,%
\footnote{%%%%%%%%%%%%%%%%%%%%%%%%%%%%%%%%%%%%%%%%%%%%%%%%%%%%%%%%%%%%%%%
This assumption can be justified by considering an adiabatic limit of the process,
namely, by replacing $X^t=\sigma^0$ in \eqref{eq:trajectry} 
with $X^t= \kappa\,\sigma^0$ and taking the limit $\kappa \to \infty$. 
This process is equivalent to the original trajectory \eqref{eq:trajectry} 
in the background with $\rmd s^2 = -\kappa^2\,\rmd t^2+\cdots$. 
The time-derivative term vanishes in the adiabatic limit $\kappa \to \infty$ 
due to the factor $H^{t\theta 3}=g^{tt}\,H_t{}^{\theta 3}$.
} %%%%%%%%%%%%%%%%%%%%%%%%%%%%%%%%%%%%%%%%%%%%%%%%%%%%%%%%%%%%%%%%%%%%%%
the equation of motion for $B_{\mu\nu}$ becomes
\begin{align}
 \partial_r \bigl(\sqrt{-g}\,\Exp{-2\phi}H^{r\theta 3}\bigr)
 = -\frac{2\kappa_4^2}{\ls^2}\, \delta(r-r_0)\,\delta(\theta-2\pi t) \,.
\end{align}
This can be integrated to obtain
\begin{align}
 \sqrt{-g}\,\Exp{-2\phi}H^{r\theta 3} = \frac{2\kappa_4^2}{\ls^2}\, \Theta(r_0-r)\,\delta(\theta-2\pi t) \,,
\label{eq:H-flux-probe}
\end{align}
where $\Theta(r)$ is the Heaviside step function. 
This non-zero field strength is produced by the probe string, 
although $B_{\mu\nu}=0$ for the original background. 
By using the explicit form of gauge fields in the KK-vortex background, 
$\cA^4= -\sigma\,(\theta/2\pi)\,\rmd x^3$ and $\cF^4= -(\sigma/2\pi)\,\rmd\theta\wedge\rmd x^3$, 
the $\FF1(4)$-charge current takes the following form:
\begin{align}
 j_4^\mu(x)
 &= -\frac{1}{2\kappa_4^2}\, \Exp{-2\phi} H^{\mu\theta 3}\, \cF^4_{\theta 3}
 -\frac{1}{2\pi\ls^2}\,\int\rmd^2\sigma \,\frac{\delta^4(x-X(\boldsymbol{\sigma}))}{\sqrt{-g}}\,\epsilon^{ab}\,V_a\, \partial_b X^\mu 
\nn\\
 &= \frac{\sigma}{2\pi\ls^2\,\sqrt{-g}}\, \Bigl[ \Theta(r_0-r)\, \delta^\mu_r 
 + \frac{\theta}{2\pi}\, \delta(r-r_0)\, \partial_0 X^\mu \Bigr]\,\delta(\theta-2\pi t) \,.
\label{eq:j_4-mu}
\end{align}
The first term represents the outflow of the $\FF1(4)$ charge to the probe string 
while the second term represents the $\FF1(4)$ charge which is localized on the probe string 
(see Figure \ref{fig:rotating-string}). 
The total amount of the $\FF1(4)$ charge localized on the probe string at $\theta$ is
\begin{align}
 q_4(\theta) = \int_V *_{\rm 4E} \mathbf{j}_4(x) = q_{\FF1}\,\frac{\theta}{2\pi} \,.
\label{eq:probe-F1}
\end{align}
Thus, after the probe goes around the KK vortex once (i.e.~$\theta=0\to 2\pi$), 
its $\FF1(4)$ charge is increased by $q_{\FF1}$, 
which agrees exactly with the expectation from the monodromy \eqref{eq:KKM-monodromy-charge}. 

We can also describe the charge transfer process from another approach 
based on a collective coordinate of the KK vortex \cite{Gregory:1997te}. 
As is discussed in \cite{Sen:1997zb}, 
the (unsmeared) KKM background has a zero-mode deformation 
associated with the $B$-field; 
$\delta \hat{B}^{(2)} = \beta\, \Omega_{\rm TN}$ 
with $\Omega_{\rm TN}$ a harmonic self-dual two form in the Taub-NUT space. 
After smearing, this zero-mode deformation takes the following form 
(see e.g.~Appendix A in \cite{Kimura:2013fda}):
\begin{align}
 \delta \hat{B}^{(2)} 
 = \beta\,\rmd \bigl[H^{-1}\,\bigl(\rmd x^4-\sigma\,(\theta/2\pi)\,\rmd x^3\bigr)\bigr] \,.
\label{eq:KKM-B-ansatz}
\end{align}
In the following, we analyze the zero-mode excitation 
by promoting $\beta$ to a dynamical variable $\beta(t)$. 
The relevant term in the action for the background fields is
\begin{align}
 -\frac{1}{2\kappa_{10}^2}\,\int \frac{1}{2}\,\hat{H}^{(3)}\wedge *_{10}\hat{H}^{(3)} 
 &=\frac{2\pi V_{3\cdots 9}}{2 \kappa_{10}^2}\,
   \int \rmd t\int_0^{r_\infty}\, \rmd r\, 
     \frac{2\pi}{r \sigma\,[\log(r_{\rm c}/r)]^3}\,\dot{\beta}(t)^2
\nn\\
 &= \frac{\sigma V_{3\cdots 9}}{4 \kappa_{10}^2\,H^2(r_\infty)}\,
    \int \rmd t \, \dot{\beta}(t)^2 \,,
\end{align}
where we have introduced an upper cutoff $r_\infty\,(\leq r_{\rm c})$ 
in the integral of radius $r$. 
We also have the following contribution from the action of the probe string:
\begin{align}
 &-\frac{1}{4\pi\ls^2}\,\int \rmd^2 \sigma \, \rmd^4 x \,\delta^4(x-X(\boldsymbol{\sigma}))\,
   \epsilon^{ab}\,\hat{B}^{(2)}_{MN}\,\partial_a X^M\,\partial_b X^N
\nn\\
 &=\frac{\sigma R_3}{2\pi\ls^2}\,\int\rmd t\,\frac{\beta(t)}{H^2(X^r)}\,
   \partial_0 \bigl[H(X^r)\,X^\theta\bigr]\, ,
\end{align}
where we have chosen 
$X^t(\boldsymbol{\sigma})=\sigma^0$, 
$X^3(\boldsymbol{\sigma})=2\pi R_3\,\sigma^1$, 
$X^r(\boldsymbol{\sigma})$ and $X^\theta(\boldsymbol{\sigma})$ arbitrary 
but $X^r(\boldsymbol{\sigma})$ is large. 
To proceed with the analysis, we need to regularize the pathological divergence 
of the background at large radius.
Here, we use an ad hoc procedure given in \cite{deBoer:2010ud}; 
we put $H(r_\infty)\sim 1$ and $H(X^r)\sim 1$ as if the background is asymptotically flat. 
Then, the equation of motion for $\beta(t)$ becomes
\begin{align}
 \ddot{\beta}(t)
 = \frac{H^2(r_\infty)}{H^2(X^r)}\, \frac{\kappa_{10}^2}{2\pi^2 \ls^2\,V_{4\cdots 9}}\,
 \partial_0\bigl[H(X^r)\,X^\theta\bigr] 
 \sim \frac{\kappa_{10}^2}{2\pi^2 \ls^2\,V_{4\cdots 9}}\,
 \partial_0 X^\theta \,,
\end{align}
and we obtain
\begin{align}
 \dot{\beta}(t) = \frac{2\kappa_{10}^2}{2\pi \ls^2\,V_{4\cdots 9}}\, \frac{X^\theta}{2\pi} \,.
\label{eq:dot-beta}
\end{align}
Thus, once the probe string goes around the center ($X^\theta=0\to 2\pi$), 
$\dot{\beta}$ becomes
\begin{align}
 \dot{\beta} = \frac{2\kappa_{10}^2}{2\pi\ls^2\,V_{4\cdots 9}} \,,
\end{align}
and, by using this value of $\dot{\beta}$, the flux integral associated with the $\FF1(4)$ charge becomes
\begin{align}
 \sigma_{\FF1(4)}^{-1}\, \int_{\partial D}\iota_9\cdots \iota_5\iota_3 *_{10} \hat{H}^{(3)} 
 =-\dot{\beta}\,\frac{2\pi\ls^2\,\sigma\,V_{35\cdots 9}}{2\kappa_{10}^2\,H(r_\infty)} 
 \sim - 1 \,,
\end{align}
where $\partial D$ is a circle with the radius $r_\infty$, 
$\iota_i$ is the interior product of the coordinate basis $\partial_i$ 
with differential forms, 
and $\sigma_{\FF1(4)}$ is a parameter defined in \eqref{eq:various-sigma} 
(see Appendix \ref{app:Page} for the Page charges of defect branes). 
That is, just a unit of $\FF1(4)$ charge is transferred 
from the background to the probe string during the process. 
Repeating the process arbitrary times, 
we can obtain a background with arbitrary number of $\FF1(4)$ charges. 

In section \ref{sec:KKM-dyon}, we explicitly construct 
a KK-vortex solution which has the same $H$-flux with that obtained from 
the $B$-field \eqref{eq:KKM-B-ansatz} (with $\dot{\beta}=\text{const.}$). 
Before that, in the next section, 
we review the notions of Alice string and Cheshire charge, 
which play important roles in the 
discussion of the charge conservation law. 

\section{Alice string and defect brane}
\label{sec:alice-defect}

In this section, we first give a detailed review of the notions of 
Alice string and Cheshire charge, 
which appear in a certain type of $(1+3)$-dimensional gauge theories.
We then argue that defect branes in string theory can be regarded as Alice strings 
and discuss the charge changing process 
in the KK-vortex background from this point of view. 

\subsection{Alice string and Cheshire charge}
\label{sec:alice-cheshire}

Alice strings \cite{Schwarz:1982ec} and Cheshire charges \cite{Alford:1990mk,Preskill:1990bm} 
appear in $(1+3)$-dimensional gauge theories 
where a charge conjugation is a gauge symmetry of the theories. 
For clearness, we explain Alice strings and Cheshire charges 
by using one of the simplest models which admits Alice strings 
\cite{Schwarz:1982ec,Alford:1990mk,Striet:2000bf}. 
Consider a $(1+3)$-dimensional gauge theory with gauge symmetry $G=SO(3)$ 
and a scalar field in the $5$-dimensional representation, 
which we denote as a real symmetric traceless $3\times 3$ matrix $\Phi_{ij}(x)$. 
By choosing a quartic potential $V(\Phi)$ with appropriate coefficients, 
the classical vacuum configuration is given by
\begin{align}
 \langle \Phi\rangle
 = \Phi_0 \equiv 
  \bpm \,v\, & 0 & 0\\
  0 & \,v\, & 0 \\
  0 & 0 & -2v\epm\,,
\end{align}
in a certain gauge. 
If we denote the generators of $G=SO(3)$ as 
$(t_k)_{ij} = -\ii\epsilon_{ijk}$ ($i,j,k=1,2,3$), 
the unbroken gauge symmetry (which keeps the vacuum configuration $\Phi_0$ invariant) 
is a subgroup $H=U(1) \rtimes \lZ_2$, 
where the $U(1)$ transformation is generated by $Q_0\equiv t_3$ 
and the $\lZ_2$ transformation is generated by
\begin{align}
 X\equiv \Exp{\ii\pi\,t_1} = 
 \bpm \,1\, & 0& 0\\
 0 & -1 & 0 \\
 0 & 0 & -1\epm\,.
\end{align}
Since $Q_0$ and $X$ satisfies $X\,Q_0\,X^{-1}=-Q_0$, 
$X$ corresponds to charge conjugation associated with the unbroken $U(1)$ gauge symmetry.

In this model, an Alice string solution can be constructed as follows. 
In order to make the energy of the configuration finite, 
we first impose $D_\mu \Phi =0$ at spatial infinity 
where $D_\mu$ is a covariant derivative associated with the $SO(3)$ gauge field. 
This condition is satisfied with
\begin{align}
 \Phi(\theta)\equiv \lim_{r\to \infty}\Phi(r,\theta)
 &= U(\theta)\, \Phi_0 \, U^{-1}(\theta) \,,
 \quad 
 U(\theta) \equiv \mathrm{P}\Exp{\ii\int_0^\theta\rmd\theta \, A_{\theta}}\rvert_{r\to \infty} \,,
\label{eq:higgs}
\end{align}
where we have set $\Phi(\theta=0) =\Phi_0$ 
and $(r,\, \theta)$ are polar coordinates in the two-dimensional space transverse to the string. 
Secondly, we impose the condition that 
$U(\theta=2\pi)$ be in the disconnected part of the gauge group;
\begin{align}
 U(\theta=2\pi) \in H_{\rm d}\,, \quad 
 H_{\rm d}\equiv \{X\,\Exp{\ii\alpha\, Q_0} \ \vert \ 0\leq \alpha<2\pi \} \,.
\end{align}
This is the defining property of Alice strings. 
Concretely, we will choose $U(\theta=2\pi)=X$, 
which can be realized by setting $A_\theta\rvert_{r\to \infty} = t_1/2$. 
This leads to the following ansatz for the global behavior of an Alice string:
\begin{align}
 A_\theta = f(r)\, t_1/2\,,\quad 
 \Phi(r,\theta) = \Exp{\ii \theta\, t_1/2} g(r) \,\Exp{-\ii \theta\, t_1/2}\,.
\end{align}
The functions $f$ and $g=(g_{ij})$ can be determined numerically 
from the equations of motion (see e.g.~\cite{Striet:2000bf}) 
although we will not need it here.

According to the $\theta$-dependence in \eqref{eq:higgs}, 
the embedding of $U(1) \rtimes \lZ_2$ into $G=SO(3)$ is non-trivially rotated around the string. 
Namely, if we define the unbroken subgroup of $G$ at $\theta$ by 
$H(\theta) \equiv \{ g \in G\ \vert \ g\, \Phi(\theta)\,g^{-1} = \Phi(\theta) \}$, 
equation \eqref{eq:higgs} leads to
\begin{align}
 H(\theta) = U(\theta)\, H(\theta=0)\, U^{-1}(\theta)\,.
\end{align}
Correspondingly, the $U(1)$ generator $Q(\theta)$ of $H(\theta)$ 
is related with the $U(1)$ generator $Q(\theta=0)$ of $H(\theta=0)$ as
\begin{align}
 Q(\theta) = U(\theta)\, Q(\theta=0)\, U^{-1}(\theta)\,.
\label{eq:Q-theta}
\end{align}
If we set $Q(\theta=0)$ to be $Q_0$, for $\theta=2\pi$, we have
\begin{align}
 Q(\theta=2\pi) = \Exp{\ii \pi t_1}\, Q_0\, \Exp{-\ii \pi t_1} = X\,Q_0\,X = -Q_0 \,.
\label{eq:flipping}
\end{align}
That is, the sign of the $U(1)$ generator becomes opposite 
as one goes around the string once.
This property is the reason why the vortex is called an Alice string; 
the monodromy works as a ``charge-conjugation looking glass'' 
\cite{Schwarz:1982ec,Bucher:1992bd}. 
Since $Q(\theta=4\pi)= Q_0$, 
the $U(1)$ generator is double-valued in the transverse two-dimensional space. 

In the presence of Alice strings, 
the electrodynamics associated with the unbroken $U(1)$ gauge group 
is called Alice electrodynamics. 
It is locally similar to the usual electrodynamics but has a strange phenomenon; 
when a charged particle goes around the Alice string once, 
the sign of the charge flips like $q\to -q$ due to property \eqref{eq:flipping}. 
Apparently this phenomenon seems to be in conflict with the charge conservation law. 

\begin{figure}[htbp]
 \centering
\includegraphics[width=70mm]{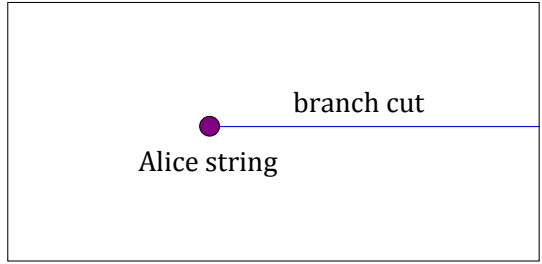}
\caption{A branch cut on the cross-sectional plane of an Alice string.} 
\label{fig:branchcut}
\end{figure}

In order to discuss the charge conservation in the Alice string background more precisely, 
we need to explain another issue closely related with the charge flipping phenomenon. 
In the presence of a single Alice string, the electric field is double-valued 
due to the double-valuedness \eqref{eq:flipping} of the $U(1)$ generator. 
If we introduce a branch cut on a half line, $\theta=0$ (see Figure \ref{fig:branchcut}),
and consider two branches which are glued together at the cut, 
the property \eqref{eq:flipping} means that the electric fields on two branches have opposite signs. 
Thus, the flux integral on a large surface is not well-defined, and we cannot define the total charge in the system.

\begin{figure}[htbp]
 \centering
 \includegraphics[width=70mm]{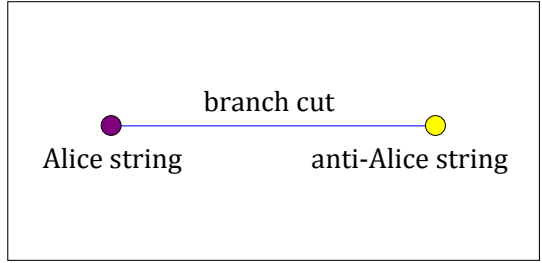}
 \caption{The cross-sectional plane of a pair of Alice and anti-Alice sting. 
 The segment between the two strings is taken as a branch cut. }
\label{fig:2alice}
\end{figure}

In contrast to an isolated string, we can define a total charge for a pair of 
an Alice string and an anti-Alice string (see Figure \ref{fig:2alice}). 
In this case, as we go around a circle which encloses the pair of strings, 
the $U(1)$ generator comes back to the original value 
since the (untraced) Wilson loop $U(2\pi)$ associated with the Alice string 
is canceled by that associated with the anti-Alice string. 
Namely, an asymptotic observer who never crosses the branch cut 
is always sitting on one side of the branch, 
and we can unambiguously define the total electric charge 
as an integral of the electric flux on the branch. 

Before we proceed further, we should comment on the definition of the electric field. 
For convenience, we take a unitary gauge in which the asymptotic value of the scalar field becomes constant 
except for on the branch cut; $\Phi=\Phi_0$. 
In this (singular) gauge, the flipping of $U(1)$ generator occurs just above the branch cut. 
Then, the charge of a particle is invariant as long as it does not cross the cut,
and its charge is flipped discontinuously only when it crosses the cut. 
We can then define the electric field at a point $x$ 
by measuring the force at $x$ felt by a test particle with charge $+q$ 
which is taken from an asymptotic region without crossing the branch cut. 
According to this definition, the electric field must change its sign at the branch cut 
since the force felt by the test particle should not change discontinuously.

\begin{figure}[htbp]
 \centering
 \includegraphics[width=110mm]{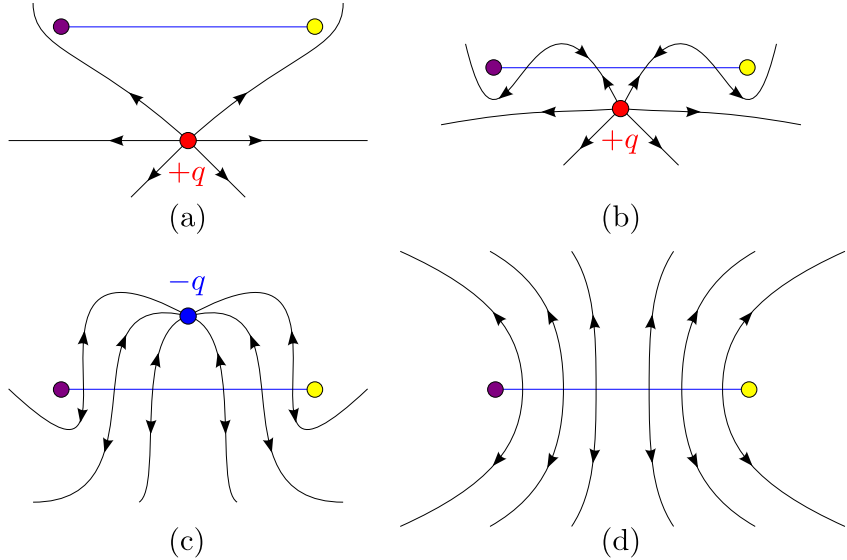}
 \caption{The process in which the Cheshire charge appears. }
 \label{fig:cheshire}
\end{figure}

Now, we can state the notion of Cheshire charge. 
A Cheshire charge is a charge without localized source 
that appears when a charged particle goes around the Alice string.
In Figure \ref{fig:cheshire}, we sketched how such a charge can appear in the process. 
In step (a) and (b), a particle with charge $+q$ is approaching the branch cut from below.
In the step from (b) to (c), the particle crossed the branch cut and the charge has become $-q$. 
After the configuration described in (c), we take the particle to infinity without crossing the cut. 
The resulting configuration is drawn in (d),
where there appears non-zero electric flux emanating from the branch cut. 
Since the branch cut is just an artifact of the singular unitary gauge, 
there is no localized source of the electric flux. 
This kind of charge (without localized source) is called a Cheshire charge. 
The total amount of the Cheshire charge can be obtained from a flux integral around the cut, 
which gives $+2q$. 
Thus, the charge conservation law is kept intact in the whole process. 

\subsection{Defect branes as Alice strings}
\label{sec:defect-alice}

Here, we explain an analogy between defect branes and Alice strings, 
by using the KK-vortex as an example of defect branes. 
Then, by using the analogy, we discuss the charge transfer process in the KK-vortex background. 

If we introduce the generalized metric for the KK-vortex background \eqref{eq:34generalized-metric}, 
it takes the following $\theta$-dependence:
\begin{align}
 &\cH^{-1}(\theta) 
 = \Omega_{\KKM}^{\rm T}(\theta) \,\cH^{-1}(\theta=0) \,\Omega_{\KKM}(\theta)\,,
\\
 &\Omega_{\KKM}(\theta) \equiv 
 \bpm 
 {\boldsymbol\omega}(\theta)^{\rm T} & \mathbf{0} \cr 
 \mathbf{0} & {\boldsymbol\omega}(\theta)^{-1} \cr 
 \epm \,,\quad 
 {\boldsymbol\omega}(\theta)
 \equiv \bpm 1& 0 \cr 
 \sigma\,(\theta/2\pi) & \,1\, \cr 
 \epm\,.
\end{align}
By using the same matrix $\Omega_{\KKM}(\theta)$, 
the charge vector of the rotating probe string at $\theta$ can be written as 
[see \eqref{eq:probe-F1}]
\begin{align}
 \vec{q}(\theta)
 \equiv \smatrix{(1/R_3)\times\#\mathrm{P}(3)\cr (1/R_4)\times\#\mathrm{P}(4)\cr (R_3/\ls^2)\times\#\mathrm{F1}(3) \cr (R_4/\ls^2)\times\#\mathrm{F1}(4)}
 = \smatrix{0\cr 0\cr (R_3/\ls^2)\times 1 \cr (R_4/\ls^2)\times \frac{\theta}{2\pi}} 
 = \Omega_{\KKM}^{-1}(\theta)\, \vec{q}(\theta=0) \,.
\label{eq:q-theta}
\end{align}
This relation is reminiscent of equation \eqref{eq:Q-theta}. 
Since the charge vector is non-trivially twisted around the center, 
the KK vortex (or general defect branes with non-trivial monodromies) 
can be regarded as Alice strings. 
Note that the discrete gauge symmetry $\lZ_2$ (in the case of the above model of Alice strings) 
is generalized to U-duality group in the case of string theory. 

As is the case with a single Alice string, 
we do not have a globally well-defined notion 
of the total winding charge in a KK-vortex background. 
However, in this case, it will be natural to define the ``total charge'' by
\begin{align}
 \vec{\mathbf{Q}} \equiv \sum_i \Omega_{\KKM}(\theta_i)\,\vec{q}_i 
 \qquad (-\infty < \theta_i<\infty) \,,
\end{align}
where $\theta_i$ is the value of the angular variable associated with the $i$-th particle. 
This definition is essentially the same as that used in \cite{deBoer:2012ma}, and, 
with this definition, the charge of the probe does not change. 
Indeed, for the case of the rotating probe string, 
the winding charge of a probe string at $\theta$ is given by
$\vec{q}(\theta) = \Omega_{\KKM}^{-1}(\theta)\, \vec{q}(\theta=0)$ [see \eqref{eq:q-theta}]
and $\Omega_{\KKM}(\theta)\,\vec{q}(\theta)$ is independent of $\theta$\,; 
i.e., no $\FF1(4)$ charge appears. 
However, in the field theoretical analysis performed in section \ref{sec:charge-transfer}, 
the probe produces a non-zero flux \eqref{eq:H-flux-probe} 
associated with the $\FF1(4)$ charge during the rotation, 
and it will be more natural to invent a definition 
such that the probe charge can change in time.

\begin{figure}[htbp]
 \centering
 \includegraphics[width=10cm]{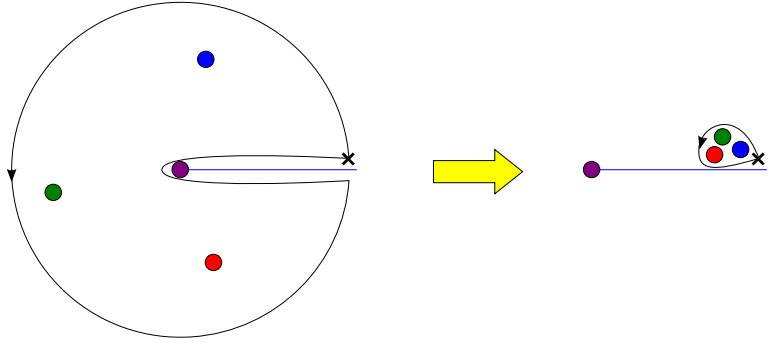}
 \caption{The conceptual procedure to define the ``total charge'' $\vec{Q}$.}
\label{fig:base-point}
\end{figure}

We thus define the ``total charge'' by
\begin{align}
 \vec{Q} \equiv \sum_i \Omega_{\KKM}(\theta_i)\,\vec{q}_i \qquad (0\leq \theta_i<2\pi)\,.
\label{fig:total-charge}
\end{align}
Conceptually, this definition can be realized by the following procedure. 
We first fix a base point on $\theta=0$ and introduce a curve which encloses all charged particles 
but does not cross the branch cut (see Figure \ref{fig:base-point}). 
Then, by shrinking the curve, we assemble all charged particles to the base point. 
Now, since all particles have the same basis for the charge vector, 
we can just add up the components of their charge vectors 
to compute the ``total charge'' $\vec{Q}$\,. 
With this definition, as is the case of $\vec{\mathbf{Q}}$\,, 
the charge remains constant during the rotation of a probe string 
as long as the string does not cross a branch cut. 
However, if the probe string crosses the branch cut, 
its winding charge must discretely change 
since the matrix discretely jump from $\Omega_{\KKM}(2\pi)$ to $\Omega_{\KKM}(0)$\,; 
\begin{align}
 &\vec{Q} = \Omega_{\KKM}(2\pi-0) \,
 \smatrix{0\cr 0\cr (R_3/\ls^2)\times 1 \cr (R_4/\ls^2)\times \frac{2\pi-0}{2\pi}} 
 = \smatrix{0\cr 0\cr (R_3/\ls^2)\times 1 \cr 0} 
\nn\\
 \to \quad 
 &\vec{Q} = \Omega_{\KKM}(+0) \,
 \smatrix{0\cr 0\cr (R_3/\ls^2)\times 1 \cr (R_4/\ls^2)\times \frac{2\pi+0}{2\pi}} 
 = \smatrix{0\cr 0\cr (R_3/\ls^2)\times 1 \cr (R_4/\ls^2)\times 1} \,.
\label{eq:jump}
\end{align}
That is, if we adopt the definition, the situation is similar to the case 
where the singular unitary gauge is taken in the presence of Alice strings, 
and the discrete change of the probe charge should be compensated by the Cheshire charge, 
which corresponds to the zero-mode deformation associated with the $B$-field. 

Geometrically, the discrete jump \eqref{eq:jump} when a probe crossed the branch cut 
can be realized as a change of coordinates given by \eqref{eq:coordinate-change}. 
This is the reason why we did not use the original coordinate $(x^3,\,x^4)$ 
but used $(x^{\prime 3},\,x^{\prime 4})$ to calculate the winding number 
in the introduction. 
For more general cases where there are multiple probe strings, 
it will be more effective to perform an active diffeomorphism 
\begin{align}
 x^{\prime 3}=x^3\,,\qquad x^{\prime 4}=x^4 + \sigma\,x^3 \,,
\end{align}
when a probe string crosses the branch cut, 
although the winding number is always measured with the use of the original coordinates. 

\section{Kaluza-Klein dyon and dyonic Kaluza-Klein vortex}
\label{sec:KKM-dyon}

In this section, we first review the construction of the KK-dyon solution 
in four-dimensional theory \cite{Roy:1998ev}. 
We point out that, if we uplift the solution to ten dimensions, 
the background has $\FF1(4)$ charges although there is no localized source, 
just like Cheshire charges. 
We then construct a KK-vortex solution with an arbitrary number of $\FF1(4)$ Cheshire charges, 
which we call a dyonic KK vortex,%%%
\footnote{As we will see later, the dyonic KK vortex is different from a usual dyon in that one of the charges of the dyonic KK vortex does not have any localized source.} 
and discuss the relation to the charge transfer process considered in section \ref{sec:charge-transfer}. 

\subsection{Kaluza-Klein dyon}

The KK-monopole solution in ten-dimensional spacetime is given by
\begin{align}
 \rmd s^2 &= - \rmd t^2+H(r)\,\bigl(\rmd r^2+r^2\,\rmd\Omega_2^2\bigr) 
    + H^{-1}(r)\,\bigl(\rmd x^4 + \omega \bigr)^2 + \rmd x^2_{56789}\,,
\\
 \Exp{2\hat{\phi}}&= 1\,,\quad 
 \hat{B}^{(2)} =0\,,\quad 
 H(r)\equiv 1+\frac{R_4}{2\,r} \,,\quad 
 \rmd\Omega_2^2 \equiv \rmd \theta^2 + \sin^2\theta \,\rmd \varphi^2\,,
\end{align}
where $R_4$ is the radius of the $x^4$-direction 
and $\omega$ satisfies $\rmd \omega = \hat{*}_{3}\,\rmd H(r)$ 
($\hat{*}_{3}$: Hodge star operator in a flat three-dimensional space), 
which can be solved by
\begin{align}
 \omega = -\frac{R_4}{2}\,(1-\cos\theta)\,\rmd\varphi\,.
\end{align}
Compactifying on the six-torus $T^6_{4\cdots 9}$ (spanned by $x^4\,,\dotsc,\,x^9$), 
we obtain the four-dimensional configuration: 
\begin{align}
 \lambda&\equiv \axion + \ii \Exp{-2\phi} = \ii\, H^{-1/2}(r) \,,
\\
 \rmd s_{\rm 4E}^2&=\Exp{-2\phi} \rmd s_4^2 
   =-H^{-1/2}(r)\,\rmd t^2+H^{1/2}(r)\,\bigl(\rmd r^2+r^2\,\rmd\Omega_2^2\bigr) \,,
\\
 \bpm \cF^4\cr \cF_4 \epm &= \bpm \rmd\cA^4\cr \rmd\cA_4 \epm\,, \qquad 
 \bpm \cA^4\cr \cA_4 \epm = \bpm \omega \cr 0 \epm \,.
\end{align}
Here, $\rmd s_{\rm 4E}^2$ is the line element in the four-dimensional Einstein frame 
while $\rmd s_4^2$ is that in the string frame. 

Now, we make a replacement $R_4\to \Delta^{1/2}\,R_4$ in the metric and $H(r)$, 
and use the $SL(2,\lR)$ duality in four dimensions (see Appendix \ref{app:ele-mag}) 
to obtain the following KK-dyon solution \cite{Roy:1998ev}:
\begin{align}
 \rmd s_{\rm 4E}^2&= -H^{-1/2}(r)\,\rmd t^2+H^{1/2}(r)\,\bigl(\rmd r^2+r^2\,\rmd\Omega_2^2\bigr) \,,
\\
 \lambda&= \axion + \ii \Exp{-2\phi} 
 =\frac{\SLa\, \SLc+\SLb\, \SLd\,H(r)}{\SLc^2+\SLd^2\,H(r)} + \ii\,\frac{H^{1/2}(r)}{\SLc^2+ \SLd^2\,H(r)} \,,
\\
 \bpm \cF^4\cr \cF_4 \epm 
 &= \bpm \rmd\cA^4\cr \rmd\cA_4 \epm\,,
\qquad 
 \bpm \cA^4\cr \cA_4 \epm
  = \bpm \SLd \, \omega \cr \frac{\SLc}{H(r)}\, \rmd t \epm \,.
\end{align}
We use the following parameterization of the $SL(2,\lR)$ matrix \cite{Roy:1998ev}:
\begin{align}
 \bpm
 \SLa & \SLb \cr
 \SLc & \SLd 
  \epm
 =
 \bpm
 \,\axion_0\, \Exp{\phi_0}\sin\alpha + \Exp{-\phi_0}\cos\alpha\, & 
 \,\axion_0\,\Exp{\phi_0} \cos\alpha - \Exp{-\phi_0} \sin\alpha\, \cr
 \Exp{\phi_0}\sin\alpha &
 \Exp{\phi_0} \cos\alpha
 \epm \,,
\end{align}
where $\phi_0$, $\axion_0$ and $\alpha$ are real free parameters to be fixed below. 
Under this $SL(2,\lR)$ transformation, the charge vector transforms as 
[see \eqref{eq:4dcharge-def} and \eqref{eq:F-G-SL2}]
\begin{align}
 \smatrix{p^4\cr p_4 \cr q^4\cr q_4}
 =
 \bpm \SLd \,\mathbf{1} & -\SLc\,\LL \cr
      -\SLb \,\LL & \SLa\,\mathbf{1} \epm 
 \smatrix{\Delta^{1/2}\,q_{\KKM}\cr 0 \cr 0 \cr 0} 
 =
 \smatrix{\SLd\,\Delta^{1/2}\,q_{\KKM}\cr 0 \cr 0 \cr -\SLb\,\Delta^{1/2}\,\gamma\,q_{\FF1}} 
\label{eq:charge-vector}
\end{align}
with $\LL = \bigl(\begin{smallmatrix} 0&1 \cr 1&0 \end{smallmatrix}\bigr)$ and 
$\gamma \equiv q_{\KKM}/q_{\FF1} = \gs^{-2}\,V_{4\cdots 9}/(2\pi\ls)^6$\,. 
We require that each component of the charge vector is an integral multiple of the elementary charge:
\begin{align}
  \SLd \,\Delta^{1/2}\equiv P \in \lZ \qquad \text{and} \qquad 
  \SLb \,\Delta^{1/2}\,\gamma \equiv Q \in \lZ \,. 
\end{align}
Using the integers $P$ and $Q$, the $SL(2,\lR)$ matrix can be written as
\begin{align}
 \bpm
 \SLa & \SLb \cr
 \SLc & \SLd 
  \epm
 &=
 \bpm
 \frac{\gamma\,P-\axion_0\,(Q-\axion_0\,\gamma\,P)}{\Delta^{1/2}\,\gamma} &
  \frac{Q}{\Delta^{1/2}\,\gamma} \cr
 -\frac{Q-\axion_0\,\gamma\,P}{\Delta^{1/2}\,\gamma} & 
 \frac{P}{\Delta^{1/2}}
 \epm \,,
\\
 \Delta^{1/2} 
 &= \sqrt{P^2 + \gamma^{-2}\,(Q-\axion_0\,\gamma\,P)^2} \,,
\end{align}
where we have set $\phi_0=0$ since the $\phi_0$ dependence 
can be absorbed into a redefinition of coordinates and the ten-dimensional dilaton. 

From the above solution, we can reconstruct the ten-dimensional solution:
\begin{align}
 \rmd s^2
  &= -\frac{\SLc^2+ \SLd^2\,H(r)}{H(r)}\,\rmd t^2
   +\bigl[\SLc^2+ \SLd^2\,H(r)\bigr]\,\bigl(\rmd r^2+r^2\,\rmd\Omega_2^2\bigr) 
\nn\\
  &\quad + H^{-1}(r)\,\bigl(\rmd x^4 + \SLd\, \omega \bigr)^2 + \rmd x^2_{56789} \,, \\
 \Exp{2\hat{\phi}} &=\frac{\SLc^2+\SLd^2\,H(r)}{H(r)}\,,\quad 
 \hat{B}^{(2)} = \frac{\SLc}{H(r)}\,\rmd t\wedge \bigl(\rmd x^4 + \SLd\, \omega\bigr)\,,
\\
 \hat{H}^{(3)} &= -\SLc\,\rmd t\wedge 
 \rmd\Bigl[\frac{1}{H(r)}\, \bigl(\rmd x^4 + \SLd\, \omega \bigr)\Bigr]\,.
\end{align}
As the charge vector \eqref{eq:charge-vector} indicates, 
this background includes $P$ KK-monopoles as well as $(-Q)$ F1(4) charge. 
The flux integral associated with F1(4) charge is calculated as
\begin{align}
 \frac{1}{2\kappa^2_{10}}\, \int_{S^2_\infty\times T^5_{5\cdots 9}}
 \Exp{-2\hat{\phi}}\,*_{10}\hat{H}^{(3)}
  = - \mu_{\FF1}\,\bigl(Q-\axion_0\,\gamma\, P\bigr)\qquad
 \Bigl(\mu_{\FF1}\equiv \frac{1}{2\pi\ls^2}\Bigr)\,,
\end{align}
where we have used $\SLc^2+\SLd^2=1$, $\SLc\,\Delta^{1/2}=-\gamma^{-1}\,(Q-\axion_0\,\gamma\,P)$, and
\begin{align}
 &\Exp{-2\hat{\phi}}\,*_{10}\hat{H}^{(3)}
\nn\\
 &=
 \left\{\begin{array}{ll}
 \rmd\Bigl[\frac{\SLc/\SLd}{\SLc^2+\SLd^2\,H(r)}\,
     \Bigl(\rmd x^4+\frac{\SLd\,\Delta^{1/2}\,R_4}{2}\,(1-\cos\theta)\,\rmd \varphi\Bigr)\Bigr]
 \wedge \rmd x^5\wedge \cdots \wedge \rmd x^9 &(\SLd\neq 0)\,,\cr
 \frac{R_4\,\Delta^{1/2}}{2}\,\sin\theta\,\rmd\theta\wedge \rmd \varphi 
 \wedge \rmd x^5\wedge \cdots \wedge \rmd x^9 & (\SLd =0)\,.
 \end{array}\right.
\end{align}
In the case of $\SLd\neq 0$, the flux integral over a sphere with the radius $r_0$ ($\times T^5_{56789}$) gives
\begin{align}
 Q_{\FF1(4)}^{0}\equiv - \mu_{\FF1}\,\frac{Q-\axion_0\,\gamma\, P}{\SLc^2+\SLd^2\,H(r_0)} \,.
\end{align}
Since the flux integral increases monotonically from $0$ to $- \mu_{\FF1}\,\bigl(Q-\axion_0\,\gamma\, P\bigr)$
as we vary $r_0$ from $0$ to $\infty$, 
we may conclude that the F1 charge is distributed over the non-compact three-dimensional space 
except the origin, $r=0$. 
However, if we consider a flux integral over an arbitrary closed surface 
which does not enclose the origin, the integral gives zero. 
Thus, it does not make sense to consider the local distribution of the source. 
We can just associate the F1 charges to a non-trivial cycle $S^2_\infty$, 
as in the case of the Cheshire charge. 

\subsection{Dyonic Kaluza-Klein vortex}
\label{sec:KKV-BG}

Here, we first construct the dyonic KK-vortex solution, and 
compute the charges both in the four-dimensional viewpoint 
and in the ten-dimensional viewpoint. 
It turns out that the background has F1(4) Cheshire charges.
We will then relate the background with the charge transfer process 
considered in section \ref{sec:charge-transfer}.

By smearing the Kaluza-Klein dyon along the $x^3$-direction, 
we obtain the following background 
(see e.g.~\cite{deBoer:2010ud,Onemli:2000kb} for the smearing procedure):%
\begin{align}
 \rmd s^2
  &= -\frac{\SLc^2+ \SLd^2\,H(r)}{H(r)}\,\rmd t^2
   +\bigl[\SLc^2+ \SLd^2\,H(r)\bigr]\,\bigl(\rmd r^2+r^2\,\rmd\theta^2+\rmd x_3^2\bigr) 
\nn\\
  &\quad + H^{-1}(r)\,\bigl(\rmd x^4 + \SLd\, \omega \bigr)^2 + \rmd x^2_{56789} \,, 
\\
 \Exp{2\hat{\phi}} &=\frac{\SLc^2+\SLd^2\,H(r)}{H(r)}\,,\qquad 
 \hat{B}^{(2)} = \frac{\SLc}{H(r)}\,\rmd t\wedge \bigl(\rmd x^4 + \SLd\, \omega\bigr)\,,
\\
 \hat{H}^{(3)} 
 &= - \SLc \,\rmd t\wedge \rmd \Bigl[ \frac{1}{H(r)}\,\bigl(\rmd x^4 + \SLd\, \omega\bigr)\Bigr]\,,
\label{eq:H3-KKd}
\end{align}
where $H(r)=(\bar{\sigma}/2\pi)\,\log(r_{\rm c}/r)$, 
$\bar{\sigma}\equiv \sigma\,\Delta^{1/2}=(R_4/R_3)\,\Delta^{1/2}$, and 
$\omega = -\bar{\sigma}\,(\theta/2\pi)\,\rmd x^3$, 
which satisfies $\rmd \omega = \hat{*}_3 \rmd H(r)$. 

The above background is a special case of the following general background with 
$\rho(z)= \ii (\bar{\sigma}/2\pi)\,\log(r_{\rm c}/z)=H(r)+\ii(\bar{\sigma}\,\theta/2\pi)$ 
and $f(z)=1$ ($z\equiv r\Exp{\ii\theta}$)\,:
\begin{align}
 \rmd s^2
  &= \bigl(\SLc^2+ \SLd^2\,\rho_2\bigr)\,
     \bigl[- \rho_2^{-1}\,\rmd t^2 + \abs{f(z)}^2\,\rmd z^2 +\rmd x_3^2\bigr] 
\nn\\
  &\quad 
 + \rho_2^{-1}\,\bigl(\rmd x^4 -\SLd\,\rho_1\,\rmd x^3\bigr)^2 + \rmd x^2_{56789} \,,
\\
 \Exp{2\hat{\phi}} &=\frac{\SLc^2+\SLd^2\,\rho_2}{\rho_2}\,,\qquad 
 \hat{B}^{(2)} = \frac{\SLc}{\rho_2}\,\rmd t\wedge 
 \bigl(\rmd x^4 -\SLd\, \rho_1\,\rmd x^3 \bigr)\,.
\end{align}
This satisfies the equations of motion 
as long as $f(z)$ and $\rho(z)\equiv \rho_1 +\ii\rho_2$ are holomorphic functions.%%%%
\footnote{%%%%%%%%%%%%%%%%%%%%%%%%%%%%%%%%%%%%%%%%%%%%%%%%%%%%%%%%%%%%%%%%%%%%%%%%%%%%
In our convention, we take the value of the dilaton at infinity to be zero. 
Thus, for a general $\rho(z)$, we need to perform a constant shift of the dilaton 
in order to satisfy this condition.
}%%%%%%%%%%%%%%%%%%%%%%%%%%%%%%%%%%%%%%%%%%%%%%%%%%%%%%%%%%%%%%%%%%%%%%%%%%%%%%%%%%%%%

\paragraph{\underline{Charges calculated in four-dimensional viewpoint}\\}

By compactifying on the six-torus $T^6_{4\cdots 9}$, we obtain the following four-dimensional configuration:
\begin{align}
 &\rmd s_{\rm 4E}^2 = - \rho_2^{-1/2}\,\rmd t^2 
     + \rho_2^{1/2}\,\bigl(\abs{f}^2\, \rmd z\,\rmd\bar{z} + \rmd x_3^2\bigr) \,,
\\
 &M =\bpm \,\rho_2\, & 0 \cr 0& \rho_2^{-1} \epm \,,\quad 
  \lambda = \axion + \ii \Exp{-2\phi}
  = \frac{\SLa\,\SLc+\SLb\,\SLd\,\rho_2}{\SLc^2+\SLd^2\,\rho_2}+\ii \frac{\rho_2^{1/2}}{\SLc^2+\SLd^2\, \rho_2} \,,
\\
 &\bpm \cF^4\cr \cF_4 \epm
  = \bpm \rmd\bigl(-\SLd\, \rho_1\,\rmd x^3\bigr) \cr 
         \rmd\bigl(\frac{\SLc\,\rmd t}{\rho_2}\bigr) \epm\,, \quad 
 \bpm \cG^4\cr \cG_4 \epm
  = \bpm \rmd\bigl(\frac{-\SLa\,\rmd t}{\rho_2}\bigr) \cr
         \rmd\bigl(\SLb\, \rho_1\,\rmd x^3\bigr) \epm\,.
\end{align}
The duality covariant charge vector is given by
\begin{align}
 \vec{Q} = 
  \smatrix{p^4\cr p_4 \cr q^4\cr q_4} 
 &= \smatrix{\frac{\SLd\, (2\pi R_3)}{2\kappa_4^2}\,\int_C\rmd\rho_1 
 \cr 0 \cr 0\cr
 -\frac{\SLb\, (2\pi R_3)}{2\kappa_4^2}\,\int_C\rmd\rho_1}
 =  \smatrix{P\,q_{\KKM} \cr 0 \cr 0 \cr -Q\,q_{\FF1}} \,,
\end{align}
where $C$ is an arbitrary closed curve on the $z$-plane which encloses the origin $z=0$ once, 
and we used $\int_C\rmd\rho_1=\bar{\sigma}$, which follows from the choice 
$\rho= \ii(\bar{\sigma}/2\pi)\,\log(r_{\rm c}/z)$. 
Thus the background contains $P$ KKM(56789,4) and $(-Q)$ F1(4) charges.

On the other hand, the physical electric charge is calculated as
\begin{align}
 \bpm Q^4\cr Q_4\epm
 &\equiv 
 \frac{1}{2\kappa_4^2}\,\bpm \int_{C_0\times S^1} \Exp{-2\phi}\,(M^{-1})^{44}\,*_{\rm 4E}\cF_4 \cr
 \int_{C_0\times S^1} \Exp{-2\phi}\,(M^{-1})_{44}\,*_{\rm 4E}\cF^4 \epm
 = \bpm 0\cr
 \frac{1}{2\kappa_4^2}\,\int_{C_0\times S^1} \frac{\SLc\,\bar{\sigma}\,\rmd(\theta/2\pi)\wedge\rmd x^3}{\SLc^2+\SLd^2\, \rho_2}\epm
\nn\\
 &=  \bpm 0 \cr -q_{\FF1}\, \frac{Q-\axion_0\,\gamma\,P}{\SLc^2+\SLd^2\,H(r_0)}\epm \,.
\end{align}
where $C_0$ is a circle with the radius $r_0$ and $S^1$ is a circle in the $x^3$-direction, 
and we used $\rho(z)=\ii (\bar{\sigma}/2\pi)\,\log(r_{\rm c}/z)$. 
The dependence on the radius $r_0$ means that the charge is distributed over the branch cut 
(although the position of the branch cut is not physical). 
If we take the limit of taking $r_0$ large and assume that $H(r_0)\to 1$ 
in the limit as in section \ref{sec:charge-transfer}, 
the physical electric charge then becomes $Q_4 = -q_{\FF1}\, (Q-\axion_0\,\gamma\,P)$\,. 

\paragraph{\underline{Charges calculated in ten-dimensional viewpoint}\\}

In the ten-dimensional viewpoint, the KKM(56789,4) charge is given by
\begin{align}
 Q_{\KKM(56789,4)} &= - \sigma_{\KKM(56789,4)}^{-1}\,
 \int_{C} \rmd \Bigl(\frac{\hat{G}_{43}}{\hat{G}_{44}} \Bigr) = P \,,
\end{align}
where we used $\int_{C} \rmd \rho_1=\bar{\sigma}_{\KKM(56789,4)}$\,.
On the other hand, by using
\begin{align}
 \Exp{-2\hat{\phi}} *_{10} \rmd\hat{H}^{(3)} 
 = \rmd \biggl[\frac{\SLc^2\,\rho_1\,\rmd x^3 + \SLd\,\rho_2\,\rmd x^4}{\SLc\,(\SLc^2+\SLd^2\, \rho_2)}\biggr] 
   \wedge \rmd x^5\wedge \cdots \wedge \rmd x^9\,,
\end{align}
we can calculate the $\FF1(4)$ charge as
\begin{align}
 \sigma_{\FF1(4)}^{-1}\,\int_C \iota_9\cdots\iota_5\iota_3 \Exp{-2\hat{\phi}}*_{10}\hat{H}^{(3)}
 = - \frac{Q-\axion_0\,\gamma\,P}{\SLc^2+\SLd^2\,H(r_0)}\,,
\label{eq:chargeonbranch}
\end{align}
where $C$ is an arbitrary curve which encloses the origin once counterclockwise 
and crosses the branch cut at $r=r_0$. 
This expression coincides with the physical electric charge in four dimensions. 

Although we can also compute the local distribution of the charges on the branch cut,\footnote{By differentiating \eqref{eq:chargeonbranch} with respect to $r_0$, we find that most charges are concentrated around the branch point $z=0$ although the charge density vanishes on $z=0$.} 
the position of the branch cut is unphysical, and so is the local charge distribution.
Thus, we can just conclude that the branch cut emanating from the KK vortex, in total, 
has $-(Q-\axion_0\,\gamma\,P)$ unit of $\FF1(4)$ charge. 
In the following, we call this kind of charges without localized source Cheshire charges. 
That is, the dyonic KK-vortex background has $-(Q-\axion_0\,\gamma\,P)$ $\FF1(4)$ Cheshire charges. 

Finally, we argue that the dyonic KK-vortex solution can be regarded as 
a resultant background after a string goes around the vortex.
To see this, we first note that the $H$-flux given in \eqref{eq:H3-KKd} is the same 
as that obtained from the $B$-field \eqref{eq:KKM-B-ansatz} with $\beta(t)=-\SLc\,\Delta^{1/2}\,t$\,. 
In section \ref{sec:charge-transfer}, we considered the case of a single KK vortex (i.e.~$P=1$). 
In order to relate the dyonic KK-vortex background with the KK-vortex background with non-zero $B$-field 
(considered in section \ref{sec:charge-transfer}), 
\eqref{eq:dot-beta} indicates that we should choose 
$-\SLc\,\Delta^{1/2} = \mu_{\FF1}\, (2\kappa_{10}^2/V_{4\cdots9})\, Q$, i.e., $\axion_0=0$. 
Thus, we will set $P=1$ and $\axion_0=0$, which makes $\SLc=- Q/(\Delta^{1/2}\,\gamma)$, $\SLd=\Delta^{-1/2}$, 
and $\Delta^{1/2}=\sqrt{1+ \gamma^{-2}\,Q^2}$\,.
With the above choice of parameters, we can interpret the dyonic KK-vortex background 
as a resultant geometry after the charge transfer process. 

\paragraph{\underline{Energy}\\}

Before closing this section, we make a brief comment on the mass of the dyonic KK vortex.
The mass can generally be computed by the following formula, 
\begin{align}
 m = \frac{1}{2\kappa_3^2}\, \int \rmd^2x \sqrt{\gamma}\,R(\gamma)\,,
\end{align}
where $\gamma$ is the transverse two-dimensional metric and 
$2\kappa_3^2\equiv 2\kappa_{10}^2/V_{3\cdots9}$.
In the three-dimensional Einstein frame, the dyonic KK-vortex solution 
has the following metric:
\begin{align}
 \rmd s^2 = - \rmd t^2 + \rho_2\,\abs{f(z)}^2\,\rmd z\,\rmd\bar{z} \,.
\end{align}
Since the dyonic deformation parameters do not appear here, 
the mass of the dyonic KK vortex is the same as that of the non-dyonic KK vortex. 

\section{$5^2_2$ background with F1 Cheshire charge}
\label{sec:522-background}

If we take a $T$-duality along the $x^3$-direction in the dyonic KK-vortex background, 
we obtain the following background:
\begin{align}
 \rmd s^2 
 &= (\SLc^2+ \SLd^2\, \rho_2)\,
 \bigl(-\rho_2^{-1}\,\rmd t^2 + \abs{f}^2\, \rmd z\,\rmd\bar{z}\bigr)
\nn\\
 &\quad + \frac{\rho_2\,\Bigl(\rmd x^3-\frac{\SLc \, \SLd \,\rho_1}{\rho_2}\,\rmd t\Bigr)^2 
 + (\SLc^2+ \SLd^2\, \rho_2)\,\rmd x_4^2}
 {\SLc^2\, \rho_2+ \SLd^2\, \abs{\rho}^2}
 + \rmd x^2_{56789} \,, 
\\
 \Exp{2\hat{\phi}} 
 &= \frac{\SLc^2+\SLd^2\,\rho_2}{\SLc^2\, \rho_2+ \SLd^2\, \abs{\rho}^2}\,,
\\
 \hat{B}^{(2)} 
 &= \frac{\SLc \, (\SLc^2+ \SLd^2\, \rho_2)}
         {\SLc^2\, \rho_2+ \SLd^2\, \abs{\rho}^2}\,\rmd t\wedge \rmd x^4
 +\frac{\SLd\, \rho_1}
       {\SLc^2\, \rho_2+ \SLd^2 \,\abs{\rho}^2}\,\rmd x^3\wedge \rmd x^4 \,,
\\
 \hat{B}^{(6)} 
 &= \frac{\SLc\,\rho_1\,\rmd x^3 + \SLd\,(\SLc^2\,\rho_2+\SLd^2\,\abs{\rho}^2)\,\rmd t}
      {\SLc^2+ \SLd^2\, \rho_2}
 \wedge \rmd x^5\wedge\cdots\wedge\rmd x^9 \,. 
\end{align}
Here, we choose the holomorphic function as 
$\rho(z)=\ii (\bar{\sigma}_{5^2_2}/2\pi)\,\log(r_{\rm c}/z)$ 
with $\bar{\sigma}_{5^2_2}\equiv \Delta^{1/2}\,\sigma_{5^2_2}$ 
and $\sigma_{5^2_2}\equiv R_3\,R_4/\ls^2$. 
If we set $\SLc=0$ and $\SLd=1$, this reproduces the standard $5^2_2$-brane background 
\cite{LozanoTellechea:2000mc,deBoer:2010ud,Kikuchi:2012za}. 
By using the generalized metric $\cH^{-1}$, we find the following relation:
\begin{align}
 &\cH^{-1}(\theta=2\pi) 
 = \Omega_{5^2_2\text{-}\FF1}^{\rm T} \, \cH^{-1}(\theta=0)\,\Omega_{5^2_2\text{-}\FF1} \,,
\\
 &\Omega_{5^2_2\text{-}\FF1} \equiv 
 \bpm \mathbf{1} & \,\mathbf{0}\, \cr
      \SLd\, \bar{\sigma}_{5^2_2}\, \boldsymbol{\varepsilon} & \mathbf{1} \epm\,, \quad 
 \boldsymbol{\varepsilon} = \bpm 0 & 1 \cr -1 & \,0\, \epm \,,
\end{align}
which follows from the discrete shift $\rho_1\to \rho_1 + \bar{\sigma}_{5^2_2}$ 
under the rotation $\theta=0\to 2\pi$. 
The monodromy matrix is exactly the same with that of the $5_2^2$ background \cite{deBoer:2010ud}, 
which we denote by $\Omega_{5^2_2}$, 
since we have chosen $\SLd\, \Delta^{1/2} =1$. 
Since the background fields are patched together with the use of a $T$-duality transformation, 
this background is a $T$-fold. 

We can compute the F1(4) charge as
\begin{align}
 \sigma_{\FF1(4)}^{-1}\,\int_C \rmd\bigl(\iota_9\cdots\iota_5\iota_3\hat{B}^{(6)}\bigr)
 = \sigma_{\FF1(4)}^{-1}\,\int_C \rmd\Bigl(\frac{\SLc\, \rho_1}{\SLc^2+ \SLd^2\, \rho_2}\Bigr) 
 = - \frac{Q}{\SLc^2+\SLd^2\,H(r_0)} \,,
\end{align}
where we used $\gamma^{-1}=\gs^{-2}\,(2\pi R_3)^2\,V_{4\cdots 9}/(2\pi\ls)^8$\,. 
This F1(4) charge becomes $-Q$ if we use the prescription $H(r_0)\to 1$\,.

In the absence of Ramond-Ramond fields, the NS5(56789) and $5^2_2(56789,34)$ charges are given by 
[see \eqref{eq:Page-NS5} and \eqref{eq:Page-522}]%%%
\footnote{See \cite{Kimura:2014wga} for a monodromy charge of a defect $(p,q)$ five-brane, which is a bound state of $p$ defect NS5-branes and $q$ exotic $5^2_2$-branes.} 
\begin{align}
 Q_{\NS5}&=\int_V *_{10}j_{\NS5} = \sigma_{{\NS}5}^{-1}\, \int_{\partial V}\rmd \hat{B}^{(2)}_{34} \,,
\\
 Q_{5^2_2}&=\int_V *_{10}j_{5^2_2}
  = \sigma_{5^2_2}^{-1}\, \int_{\partial V}
 \rmd \biggl(\frac{\hat{B}^{(2)}_{34}}{\det(\hat{G}_{ab}+\hat{B}^{(2)}_{ab})} \biggr)\,.
\end{align}
In the above background, these charges become
\begin{align}
 Q_{\NS5} = \sigma_{{\NS}5}^{-1}\, 
   \int_{\partial V}\rmd \Bigl[\frac{\SLd\, \rho_1}{\SLc^2\, \rho_2+ \SLd^2 \,\abs{\rho}^2}\Bigr] \,,\quad 
 Q_{5^2_2} = \SLd\,\sigma_{5^2_2}^{-1}\, 
   \int_{\partial V} \rmd \rho_1 = 1 \,,
\end{align}
where $\partial V$ is a closed curve which encloses the center once counterclockwise, 
and we used $\rho=\ii(\bar{\sigma}_{5^2_2}/2\pi)\,\log(r_{\rm c}/z)$ for the latter expression. 
Since $\abs{\rho}^2=(\bar{\sigma}_{5^2_2}/2\pi)^2\,\bigl[\log^2(r_{\rm c}/z)+\theta^2\bigr]$ 
is a multi-valued function, 
the integral in $Q_{\NS5}$ along the closed curve $\partial V$ depends on the choice of the starting point 
(i.e.~if we choose $\partial V$ as a circle, 
$\int_{\partial V} =\int_{\theta}^{\theta+2\pi}$, the integral depends on $\theta$) 
and we cannot define a meaningful charge for NS5-brane \cite{deBoer:2012ma}. 
This kind of peculiar charge also appears in other duality frames. 
For example, in the D7 background \eqref{eq:D7-background}, which we will show later, 
there apparently exist non-zero NS7 charges 
$Q_{\NS7}\propto \int_C \rmd \bigl(\hat{C}^{(0)}/\abs{\hat{C}^{(0)}+\ii\Exp{-\hat{\phi}}}^2\bigr)$. 
However, it does not have a meaningful value in the same reasoning with the above NS5 charge 
(see e.g.~section 5 of \cite{Meessen:1998qm} for discussions of seven-brane charges from other viewpoints). 

\subsection{Winding process in the ``doubled'' $5^2_2$ background}

The above background can be interpreted as a resultant background after 
a probe with $\PP(3)$ charge has rotated around the center of the $5^2_2$ background. 
Indeed, as a probe moves around the center once, 
its charge changes by the action of the $T$-duality monodromy as
\begin{align}
 \smatrix{(1/R_3)\times\#\mathrm{P}(3)\cr (1/R_4)\times\#\mathrm{P}(4)\cr
  (R_3/\ls^2)\times\#\mathrm{F1}(3) \cr (R_4/\ls^2)\times\#\mathrm{F1}(4)}
 = \smatrix{(1/R_3)\times 1\cr 0\cr 0 \cr 0}
 \to 
   \Omega_{5^2_2}^{-1}\,
 \smatrix{(1/R_3)\times 1\cr 0\cr 0 \cr 0}
 = \smatrix{(1/R_3)\times 1\cr 0\cr 0 \cr (R_4/\ls^2)\times 1} \,, 
\label{eq:522charge-change}
\end{align}
and the charge of the probe brane becomes $\mathrm{P}(3) + \FF1(4)$. 
The additional $\FF1(4)$ charge should be compensated by the background flux 
as it was the case in the T-dual frame (i.e.~in the KK-vortex frame).
We can again perform a similar analysis for the $5^2_2$ background. 
However, it is rather difficult to understand intuitively 
how a probe string with just a momentum charge can be converted into a winding string. 
In order to understand the winding process geometrically, 
it will be beneficial to describe the $5^2_2$ background in terms of a doubled geometry, 
which was found in the study of string field theory \cite{Kugo:1992md} and the double field theory 
\cite{Hull:2009mi,Hull:2009zb,Hohm:2010jy,Hohm:2010pp,Hohm:2011dv,Hull:2006va} 
(see also \cite{Geissbuhler:2013uka,Aldazabal:2013sca,Hohm:2013bwa} for reviews of the double field theory). 

Let us decompose the ten-dimensional local coordinates as $(x^M)=(x^\mu,\,x^a)$, 
where $x^\mu$ are the coordinates in the non-compact spacetime $(\mu =0,1,\dotsc,9-d)$ 
and $x^a$ are the coordinates on the $d$-torus $(a =10-d,\dotsc,9)$. 
In the double field theory, in addition to the coordinates $x^a$ 
(associated with momentum excitations $p_a=-\ii\partial_a$), 
we introduce new (periodic) coordinates, $\tilde{x}_a$ 
(associated with winding excitations $w^a=-\ii\tilde{\partial}^a\equiv -\ii\partial/\partial \tilde{x}_a$), 
and deal with these coordinates $(x^I)\equiv (\tilde{x}_a,\,x^a)$ on an equal footing. 
With these $(10 + d)$ coordinates $(x^A)\equiv (x^\mu,\,x^I)$, 
the low energy effective theory becomes manifestly covariant under the T-duality $O(d,d)$ transformations, 
and the original gauge symmetries in the NS-NS fields 
(i.e.~diffeomorphisms on the $d$-torus and gauge transformations associated with the $B$-field) 
are included in the symmetry. 
As is shown in \cite{Hohm:2012gk}, under the generalized coordinate transformation, 
a generalized tensor transforms as
\begin{equation}
 V'_I(x')=\mathcal{F}_I{}^J\, V_J(x)\,, 
\end{equation}
where the matrix $\mathcal{F}\equiv (\mathcal{F}_I{}^J)$ is given by
\begin{equation}
 \mathcal{F}_I{}^J 
 \equiv \frac{1}{2}\,\Bigl(\frac{\partial x^K}{\partial x^{\prime I}}\frac{\partial x'_K}{\partial x_J}
        +\frac{\partial x'_I}{\partial x_K}\frac{\partial x^J}{\partial x^{\prime K}}\Bigr)\,,\quad 
 (x_I)\equiv \bigl(\eta_{IJ}\,x^J\bigr) = (x^a,\,\tilde{x}_a)\,.
\end{equation}
Since the generalized metric $\cH^{-1}$ behaves as a tensor under the generalized coordinate transformations, 
we have
\begin{align}
 \cH'^{-1} = \mathcal{F}\,\cH^{-1}\,\mathcal{F}^{\rm T} \,,\quad 
 \cH^{-1}\equiv 
 \bpm 
  \hat{G}^{-1} & -\hat{G}^{-1}\,\hat{B} \cr
  \hat{B}\, \hat{G}^{-1} & \hat{G}-\hat{B}\, \hat{G}^{-1}\,\hat{B}
 \epm \,.
\end{align}

Now, we go back to the case of the $5^2_2(56789,34)$ background. 
In this case, associated with the 3-4 torus, 
we introduce two winding coordinates $(\tilde{x}_3,\,\tilde{x}_4)$ with periods $(2\pi\ls^2/R_3,\,2\pi\ls^2/R_4)$ 
and consider the 12-dimensional background. 
The $T$-duality monodromy of the $5^2_2$ background can be realized as 
the following generalized coordinate transformation:
\begin{align}
 x^{\prime 4} = x^4 + \sigma_{5^2_2}\, \tilde{x}_3\,,\quad 
 x^{\prime 3} = x^3 \,,\quad \tilde{x}'_a=\tilde{x}_a \,.
\end{align}
Now, let us consider a probe string with a $\PP(3)$ charge. 
Just like a probe string with a winding charge, $\FF1(3)$, wraps around the $x^3$-direction once, 
a string with a momentum charge, $\PP(3)$, wraps around the $\tilde{x}_3$-direction once. 
Then, the above coordinate transformation implies that, 
after the string with a $\PP(3)$ charge goes around the center of the $5^2_2$ background, 
it gets an additional winding charge in the $x^{\prime 4}$-direction. 
Since the length of the string in the $x^{\prime 4}$-direction is given by 
$\sigma_{5_2^2} \times 2\pi\ls^2/R_3 = 2\pi R_4$, 
the string wraps around the direction exactly once. 
In this way, the charge change \eqref{eq:522charge-change} can be 
understood geometrically even in the non-geometric $5^2_2$ background. 

This is quite similar to the case of the rotating string in the KK-vortex background 
discussed in the introduction. 
If there is a more general formulation in which the $U$-dualities 
can be realized as some kind of diffeomorphism, 
this kind of charge transfer process can be understood geometrically 
in any duality frame. 

\section{Alice string backgrounds with Cheshire charges}
\label{sec:various-alice}

We can construct various Alice string backgrounds by taking the following dualities:
\begin{align*}
 &\smatrix{\mathbf{KKM(356789,4)}_{\rm Alice}\cr \mathbf{F1(4)}_{\rm Cheshire}}
 \overset{T_3}{\longrightarrow}
  \smatrix{\mathbf{5^2_2(56789,34)}_{\rm Alice}\cr \mathbf{F1(4)}_{\rm Cheshire}}
 \overset{S^*}{\longrightarrow}
  \smatrix{\mathbf{5^2_3(56789,34)}_{\rm Alice}\cr \mathbf{D1(4)}_{\rm Cheshire}}
\\
 \overset{T_4}{\longrightarrow}
 &\smatrix{\mathbf{6^1_3(456789,3)}_{\rm Alice}\cr \mathbf{D0}_{\rm Cheshire}}
 \overset{T_3}{\longrightarrow}
 \smatrix{\mathbf{NS7(3456789)}_{\rm Alice}\cr \mathbf{D1(3)}_{\rm Cheshire}}
 \overset{S^*}{\longrightarrow}
 \smatrix{\mathbf{D7(3456789)}_{\rm Alice}\cr \mathbf{F1(3)}_{\rm Cheshire}}
\\
 \overset{T_{4\cdots9}}{\longrightarrow}
 &\smatrix{\mathbf{D1(3)}_{\rm Alice}\cr \mathbf{F1(3)}_{\rm Cheshire} ?}
 \overset{S}{\longrightarrow}
 \smatrix{\mathbf{F1(3)}_{\rm Alice}\cr \mathbf{D1(3)}_{\rm Cheshire} ?}
 \overset{T_{456789}}{\longrightarrow}
 \smatrix{\mathbf{F1(3)}_{\rm Alice}\cr \mathbf{D7(3456789)}_{\rm Cheshire} ?} \,,
\end{align*}
where $S^*$ represents the inverse of the $S$-duality transformation.

In the following, we construct various Alice string backgrounds 
and show that the Cheshire charges disappear in the last three duality frames. 
In addition, we show that these backgrounds can be obtained from 
a pure F1 background by a sequence of $T$-dualities and $SL(2)$ dualities in type IIB theory.

\paragraph{\underline{D7 Alice + F1 Cheshire charge}\\}

By the above chain of dualities, we obtain the D7 background with F1(3) Cheshire charge:
\begin{align}
 \rmd s^2
 &= (\SLc^2+\SLd^2\,\rho_2)^{1/2} \,
    \bigl[\,\rho_2^{-1}\,(-\rmd t^2 + \rmd x_3^2) + \abs{f}^2\,\rmd z\,\rmd\bar{z} \,\bigr] 
 + (\SLc^2+\SLd^2\,\rho_2)^{-1/2}\, \rmd x^2_{456789} \,,
\\
 \Exp{2\hat{\phi}} 
 &= \frac{1}{\rho_2\,(\SLc^2+\SLd^2\,\rho_2)} \,,\quad 
 \hat{B}^{(2)} 
  = -\frac{\SLc}{\rho_2}\, \rmd t \wedge \rmd x^3 \,,\quad
 \hat{B}^{(6)} 
  =\frac{\SLc\,\rho_1}{\SLc^2+\SLd^2\,\rho_2}\,\rmd x^4\wedge\cdots\wedge\rmd x^9\,,
\\
 \hat{C}^{(0)} &= \SLd\,\rho_1 \,,\quad 
 \hat{C}^{(2)} = \frac{\SLc\,\SLd\,\rho_1}{\rho_2}\,\rmd t\wedge \rmd x^3 \,,\quad 
 \hat{C}^{(4)} =0 \,.
\label{eq:D7-background}
\end{align}
The D7-brane charge is given by
\begin{equation}
 Q_{\DD7} = \sigma_{\DD7}^{-1}\, \int_{C}\rmd \hat{C}^{(0)} = 1 \qquad 
 (\sigma_{\DD7}\equiv \gs)\,,
\end{equation}
where we used $\SLd\,\Delta^{1/2} =1$\,.
Note that the monodromy of the background is exactly the same as that of the pure D7 background:
\begin{align}
 \tau(z) \equiv \hat{\cC}^{(0)}+\ii\Exp{-\hat{\Phi}} \to \tau(z) + 1 \,,\quad
 \bpm \hat{\cC}^{(2)} \cr \hat{B}^{(2)}\epm
 \to \bpm \,1\, & -1 \cr 0 & 1\epm 
   \bpm \hat{\cC}^{(2)} \cr \hat{B}^{(2)}\epm \,,
\end{align}
where $\hat{\cC}^{(p)}\equiv \gs^{-1}\,\hat{C}^{(p)}$\,. 
In addition, we can calculate the F1(3) Cheshire charge as
\begin{align}
 \sigma_{\FF1(3)}^{-1}\, \int_{C} \rmd\Bigl[\frac{\SLc\,\rho_1}{\SLc^2+\SLd^2\,\rho_2}\Bigr]
 = - Q \,,
\end{align}
where we used $\SLc^2+\SLd^2\,H(r_0)=1$\,.
Note that although there is non-zero $\hat{C}^{(2)}$, 
there are no D5 Page charges since $\hat{C}^{(2)}+\hat{B}^{(2)}\,\hat{C}^{(0)}=0$\,.

\paragraph{\underline{D1 Alice + ``F1 Cheshire charge''}\\}

By further taking $T_{456789}$-duality, we obtain
\begin{align}
 \rmd s^2
 &= (\SLc^2+\SLd^2\,\rho_2)^{1/2} \,
    \bigl[\,\rho_2^{-1}\,(-\rmd t^2 + \rmd x_3^2) + \abs{f}^2\,\rmd z\,\rmd\bar{z} 
 + \rmd x^2_{456789}\,\bigr] 
 \,,
\\
 \Exp{2\hat{\phi}} &= \frac{(\SLc^2+\SLd^2\,\rho_2)^2}{\rho_2} \,,\quad 
 \hat{B}^{(2)} = -\frac{\SLc}{\rho_2}\, \rmd t \wedge \rmd x^3 \,,
\\
 \hat{C}^{(0)} &= \frac{\SLc}{\SLd\,(\SLc^2+\SLd^2\,\rho_2)} \,,\quad 
 \hat{C}^{(2)} = \frac{1}{\SLd\,\rho_2}\,\rmd t\wedge \rmd x^3 \,, \quad 
 \hat{C}^{(4)} = 0 \,,
\\
 \hat{C}^{(6)} &= \SLd\,\rho_1\,\rmd x^4\wedge \cdots \wedge \rmd x^9 \,,\quad 
 \hat{C}^{(8)} = \frac{\SLc\,\SLd\,\rho_1}{\rho_2}\,\rmd t\wedge \rmd x^3 \wedge \cdots \wedge \rmd x^9 \,.
\end{align}
In fact, this background can be obtained by acting 
the $SL(2)$ transformation (in type IIB theory) to the pure D1(3) background 
(i.e.~the above background with $\SLc=0$ and $\SLd=1$):
\begin{align}
 \tau = \frac{\SLd^{-1}\,(\ii\gs^{-1}\,\rho_2^{-1/2}) + 0}{\SLc\,\gs\,(\ii\gs^{-1}\,\rho_2^{-1/2}) + \SLd}\,,\quad
  \bpm \hat{\cC}^{(2)} \cr \hat{B}^{(2)}\epm
  =\bpm \SLd^{-1} & 0\cr -\SLc\,\gs & \,\SLd\,\epm 
   \bpm \frac{1}{\gs\rho_2}\rmd t\wedge \rmd x^3 \cr 0\epm \,.
\end{align}
Strangely enough, in this frame, there are no expected F1(4) Cheshire charges. 
Indeed, we have 
\begin{align}
 \hat{B}^{(6)} + \hat{C}^{(4)} \wedge \hat{C}^{(2)} 
 - \frac{\hat{C}^{(0)}\,\hat{C}^{(2)} \wedge\hat{C}^{(2)} \wedge\hat{C}^{(2)}}{6\,\abs{\tau}^2} =0 \,,
\end{align}
and, from \eqref{eq:PageF1}, we have $\int_V *_{10}j_{\FF1}=0$ for any region $V$.

\paragraph{\underline{F1 Alice + ``D1 Cheshire charge''}\\}

By taking $S$-duality, we obtain
\begin{align}
 \rmd s^2
 &= \SLd^{-1}\,\bigl[\,\rho_2^{-1}\,(-\rmd t^2 + \rmd x_3^2) + \abs{f}^2\,\rmd z\,\rmd\bar{z} 
 + \rmd x^2_{456789}\,\bigr] 
 \,,
\\
 \Exp{2\hat{\phi}} &= \frac{1}{\SLd^4\,\rho_2} \,,\quad 
 \hat{B}^{(2)} = -\frac{1}{\SLd\,\rho_2}\,\rmd t\wedge \rmd x^3 \,,
\quad
 \hat{B}^{(6)} = \SLd\,\rho_1\,\rmd x^4\wedge \cdots\wedge \rmd x^9 \,,
\\
 \hat{C}^{(0)} &= -\SLc\,\SLd\,,\quad 
 \hat{C}^{(2)} 
  = -\frac{\SLc}{\rho_2}\, \rmd t \wedge \rmd x^3\,,\quad 
  \hat{C}^{(p)}=0 \quad (p=4,6,8) \,.
\label{eq:F1-D1}
\end{align}
As in the case of the above D1 background, 
this background also does not have Cheshire charge. 
In addition, we note that this background can be obtained 
by acting the $SL(2)$ transformation (in type IIB theory) to the F1(3) background;
\begin{align}
 \tau =\frac{\SLd\,(\ii\gs^{-1}\rho_2^{1/2}) - \SLc\,\gs^{-1}}{0 + \SLd^{-1}}\,,\quad
  \bpm \hat{\cC}^{(2)} \cr \hat{B}^{(2)}\epm
  =\bpm \,\SLd\, & \SLc\,\gs^{-1} \cr 0 & \SLd^{-1} \epm 
   \bpm 0 \cr -\frac{1}{\rho_2}\,\rmd t\wedge \rmd x^3\epm \,.
\label{eq:F1-SL2}
\end{align}
In section \ref{sec:KKM-dyon}, we constructed the dyonic KK-vortex 
by using the electric-magnetic duality in four-dimensional theory. 
However, as we found here, we can construct the background 
only from the ten-dimensional point of view, 
without considering the compactification to four dimensions. 

\paragraph{\underline{F1 background}\\}

By taking $T_{456789}$-duality, we obtain
\begin{align}
 \rmd s^2
 &= \SLd^{-1}\,\bigl[\,\rho_2^{-1}\,(-\rmd t^2 + \rmd x_3^2) + \abs{f}^2\,\rmd z\,\rmd\bar{z}\, \bigr] 
 + \SLd\, \rmd x^2_{456789}
 \,,
\\
 \Exp{2\hat{\phi}} 
 &= \frac{\SLd^2}{\rho_2} \,,\quad 
 \hat{B}^{(2)} = -\frac{1}{\SLd\,\rho_2}\,\rmd t\wedge \rmd x^3 \,,
\quad
 \hat{B}^{(6)} = \SLd\,\rho_1\,\rmd x^4\wedge \cdots\wedge \rmd x^9 \,,
\\
 \hat{C}^{(0)} &=0\,, \quad \hat{C}^{(2)} =0\,, 
 \quad \hat{C}^{(4)} =0\,,
\\
 \hat{C}^{(6)} &= -\SLc\,\SLd\,\rmd x^4\wedge \cdots\wedge \rmd x^9 \,,\quad 
 \hat{C}^{(8)} 
  = -\frac{\SLc}{\rho_2}\,\rmd t \wedge \rmd x^3\wedge \cdots\wedge \rmd x^9\,. 
\label{eq:F1-D7}
\end{align}
Note that all the standard (i.e.~$p\leq 4$) Ramond-Ramond potentials $\hat{C}^{(p)}$ are zero. 
Thus, we conclude that this background is exactly the same as the F1 background. 

Indeed, in the democratic formulation of supergravity 
\cite{Fukuma:1999jt,Bergshoeff:2001pv}, 
there is a gauge invariance under 
(see e.g.~Appendix A in \cite{Kimura:2014upa}, which uses the same conventions)
\begin{align}
 \delta \hat{C}^{(6)} = \rmd \Lambda^{(5)}\,,\quad 
 \delta \hat{C}^{(8)} = -\hat{B}^{(2)}\wedge \rmd \Lambda^{(5)}\,.
\end{align}
Using this gauge transformation with 
$\rmd\Lambda^{(5)}=\SLc\,\SLd\,\rmd x^4\wedge \cdots\wedge \rmd x^9$, 
we can totally remove the Ramond-Ramond potentials. 
After the transformation, we can obtain the standard F1 background 
by making the shift of the dilaton and rescaling of coordinates. 

\section{Conclusion and discussions}
\label{sec:conclusion}

Among various objects in string theory, 
defect branes have distinguished features that 
their background geometries have non-trivial monodromies.
In this paper, we examined a similarity between 
defect branes in string theory and Alice strings in four-dimensional gauge theories. 
By using the analogy, U-duality monodromies of defect-brane backgrounds 
can be thought of as a generalization 
of the charge-conjugation $\lZ_2$-transformation of Alice strings. 
In order to develop the analogy further, we first examined a rotating probe brane in a defect-brane background 
and showed that the missing charge of the probe brane is transferred into the background 
and thus the whole amount of charge is conserved during the process. 

We then explicitly constructed a dyonic defect-brane solution, 
i.e.~a defect-brane background with additional charges 
that are the same as the charges transferred from the probe brane. 
Curiously enough, it turns out that the additional charges have no localized sources, 
and so they can be regarded as Cheshire charges. 
In this way, we argued that the dyonic defect-brane background corresponds to 
the resulting configuration after the probe brane goes around the defect. 

Though we shed light on puzzling aspects of defect branes in string theory in this paper,
there are many directions that remain to be explored further. 
These include the following problems. 
The authors hope to report these issues in the near future.

\begin{itemize}
\item As we showed in section \ref{sec:KKM-dyon}, 
the unsmeared KK-dyon background does have a Cheshire-like charge, 
namely, a charge with no localized source. 
While Cheshire charges are generally associated with a branch cut 
which extends between Alice strings, in the KK-dyon background, 
the fields are single-valued and there is no branch cut. 
In this case, as discussed in \cite{Gregory:1997te}, 
the Cheshire-like charge appears as a result of the unwinding of a probe string in the KKM background,
which is caused by a coordinate singularity of $S^3$,
\begin{align}
 \rmd\ell^2_{S^3}
 =\bigl[\rmd x^4+(R_4/2)\,(1-\cos\theta)\,\rmd\varphi\bigr]^2 +
r^2\,\rmd \Omega_2^2\,,
\end{align}
at the south pole, $\theta=\pi$. 
It will be interesting to explore a topological understanding further 
for the Cheshire-like charge in the KK-dyon solution 
as in the case of a Cheshire charge in an Alice string background \cite{Bucher:1992bd}. 

\item In section \ref{sec:various-alice}, by taking dualities, 
we obtained the duality frames where a Cheshire charge is absent. 
Since the existence of the Cheshire charge in the original KK-vortex background 
is related to the existence of the zero-mode deformation associated with the $B$-field, 
the disappearance of the Cheshire charge may imply that 
there are no zero-mode deformations in the D1 or F1 background 
associated with $\hat{C}^{(2)}$ or the $B$-field. 
It would be interesting to study this point further. 
Moreover, by further taking dualities, 
we obtained an F1 background which has 
non-zero Ramond-Ramond $p$-form with $p>5$ [see \eqref{eq:F1-D7}]. 
We found that this background can be obtained from the pure F1-background 
by the action of a gauge transformation associated 
with the 5-form gauge parameter $\Lambda^{(5)}$\,. 
On the other hand, the background \eqref{eq:F1-D1}, 
which is $T_{456789}$-dual to the background \eqref{eq:F1-D7}, 
can also be obtained by acting an $SL(2)$ duality \eqref{eq:F1-SL2} 
to the pure F1-background. 
In this sense, the $SL(2)$-duality transformation 
can be realized a combination of the $T$-dualities and the gauge transformations 
associated with the Ramond-Ramond fields of higher degree. 
It will be interesting to study the relation between 
the general $SL(2)$-duality transformation in type IIB theory 
and the gauge transformations of Ramond-Ramond fields 
and $T$-dualities further. 

\item In order to understand the charge changing process in defect-brane backgrounds 
in a completely geometrical way, 
a duality covariant formulation for supergravity and probe action is desired. 
For example, as discussed in section \ref{sec:522-background}, 
the double field theory gives an intuitive interpretation 
of the charge changing process in the $5^2_2$ background. 
However, to obtain a complete description of the dynamics, 
a T-duality covariant string action, that is, so-called the double sigma model 
\cite{Tseytlin:1990hn,Tseytlin:1990va,Tseytlin:1990nb,Copland:2011wx} is needed. 
It is interesting to explore more on how the charge changing process 
can be described within the double sigma model. 
Further, it would be much more advantageous to reformulate the low energy supergravity theory 
in a fully U-duality covariant manner 
(see, for example, 
\cite{Hohm:2013jma,Blair:2013gqa,Hohm:2013vpa,Hohm:2013uia,Park:2014una,Godazgar:2014nqa,Hohm:2014fxa} 
for recent works) 
and realize the U-duality transformations as geometric transformations, 
like the generalized diffeomorphism in the double field theory. 
In such formulation, we can geometrically describe the charge transfer process 
in any defect-brane background (e.g.~$\DD1\to \DD1+\FF1$ process in the D7 background). 
It would be also interesting to apply such a U-duality covariant formulation 
for other charge changing process in string theory, such as the Hanany-Witten transition \cite{Hanany:1996ie}.

\item In section \ref{sec:various-alice}, 
we found that the defect-brane backgrounds with Cheshire charges can be constructed 
not from the four-dimensional electric-magnetic duality 
but more directly from dualities in ten-dimensional supergravity. 
By using the ten-dimensional dualities, 
we can easily construct various supertube solutions with (dipole) Cheshire charges, 
e.g., from the F1-P system considered in \cite{deBoer:2012ma}; 
the explicit form of these backgrounds and the analysis of their properties will be given elsewhere. 
It will be more interesting to construct a supertube solution with Cheshire charges 
which can be observed at infinity but does not have localized source, 
and examine their roles as black hole microstates in the fuzzball program 
\cite{Mathur:2005zp,Bena:2007kg,Skenderis:2008qn,Balasubramanian:2008da}. 
\end{itemize}

\section*{Acknowledgments}

We acknowledge support by the National Research Foundation of Korea (NRF-MSIP) 
grants 2005-0093843, 2010-220-C00003 and 2012-K2A1A9055280. 
We would like to thank David Andriot, Andr\'e Betz, Toru Kikuchi, 
Tetsuji Kimura, Soo-Jong Rey, Shin Sasaki, and Masaki Shigemori 
for helpful discussions. 

\appendix

\section{Conventions and notations}
\label{app:conventions}

We denote the degree of a $p$-form as $\alpha^{(p)}$, 
and use the following definition of the Hodge dual:
\begin{align}
 *_d (\rmd x^{\mu_1}\wedge\cdots\wedge \rmd x^{\mu_p})
 &\equiv \frac{1}{(d-p)!}\,\varepsilon^{\mu_1\cdots\mu_p}{}_{\nu_1\cdots\nu_{d-p}}\,
 \rmd x^{\nu_1}\wedge\cdots \wedge \rmd x^{\nu_{d-p}}\,,
\\
 (*_d\alpha^{(p)})_{\mu_1\cdots\mu_{d-p}}
 &= \frac{1}{p!}\,\varepsilon^{\nu_{1}\cdots\nu_{p}}{}_{\mu_1\cdots\mu_{d-p}}\,
    \alpha_{\nu_1\cdots\nu_p}
\end{align}
with $\varepsilon^{01\cdots (d-1)}=-1/\sqrt{-g}$ 
and $\varepsilon_{01\cdots (d-1)}= +\sqrt{-g}$\,.
Then, the following relation follows:
\begin{equation}
 \alpha^{(p)} \wedge *_d\beta^{(p)}
 = \beta^{(p)} \wedge *_d\alpha^{(p)}
 = \frac{1}{p!}\,\alpha_{\mu_1 \cdots \mu_p}\,\beta^{\mu_1 \cdots \mu_p}\,\sqrt{-g}\,\rmd^dx \,.
\end{equation}

We denote various charges of branes in string theory in the following way. 
For a D$p$-brane which is (spatially) extending in the $x^{i_1},\dotsc,x^{i_p}$-directions, 
its charge is denoted by ${\DD}p(i_1\dotsc i_p)$. 
The winding and momentum charge of a string are also denoted in the same way; $\FF1(i)$ and $\PP(i)$. 
For various defect branes which are extending or smeared in the seven-torus $T_{3456789}$ 
and have the following mass, the charge is denoted by $b_n^c(i_1\cdots i_b,j_1\cdots j_c)$:
\begin{align}
 M = \frac{R_{i_1}\cdots R_{i_b}\,(R_{j_1} \cdots R_{j_c})^2}{\gs^n\, \ls^{b+2c+1}}\,,
\end{align}
where $R_{i}$ is the compactification radius in the $x^i$-direction and $\gs$ is the string coupling constant. 
In particular, for well-known branes, we denote ${\NS5}(i_1\cdots i_5)\equiv 5_2(i_1\cdots i_5)$, 
${\rm KKM}(i_1\cdots i_5,j)\equiv 5_2^1(i_1\cdots i_5,j)$, and ${\NS7}(3\cdots 9)\equiv 7_3(3\cdots 9)$. 
See \cite{deBoer:2012ma} for greater detail. 

\subsection{Supergravity actions and duality rules}

In our convention, the action for type II supergravity is given by
\begin{align}
 S_{\rm IIA} 
 &= \frac{1}{2\kappa_{10}^2}\,
    \int \Exp{-2\hat{\phi}} \bigl(*_{10} \hat{R} + 4\,\rmd\hat{\phi}\wedge *_{10}\rmd\hat{\phi}
          - \frac{1}{2}\,\hat{H}^{(3)}\wedge *_{10}\hat{H}^{(3)}\bigr)
\nn\\
 &\quad -\frac{1}{4\kappa_{10}^2}\,
    \int \bigl(\hat{G}^{(2)}\wedge *_{10}\hat{G}^{(2)}
    + \hat{G}^{(4)}\wedge *_{10}\hat{G}^{(4)} 
    - \hat{B}^{(2)}\wedge \rmd \hat{C}^{(3)}\wedge \rmd \hat{C}^{(3)}\bigr)\,,
\\
 S_{\rm IIB} 
 &= \frac{1}{2\kappa_{10}^2}\,
    \int \Exp{-2\hat{\phi}} \bigl(*_{10} \hat{R} + 4\,\rmd\hat{\phi}\wedge *_{10}\rmd\hat{\phi} 
     - \frac{1}{2}\,\hat{H}^{(3)}\wedge *_{10}\hat{H}^{(3)}\bigr)
\nn\\
 &\quad -\frac{1}{4\kappa_{10}^2}\,
    \int \bigl(\hat{G}^{(1)}\wedge *_{10}\hat{G}^{(1)}
    + \hat{G}^{(3)}\wedge *_{10}\hat{G}^{(3)} 
    + (1/2)\,\hat{G}^{(5)}\wedge *_{10}\hat{G}^{(5)} \bigr)
\nn\\
 &\quad - \frac{1}{4\kappa_{10}^2}\,\int \Bigl(\hat{C}^{(4)}+ \frac{1}{2}\, \hat{B}^{(2)}\wedge \hat{C}^{(2)}\Bigr)
          \wedge \rmd \hat{C}^{(2)}\wedge \hat{H}^{(3)}\,,
\end{align}
where $2\kappa_{10}^2 \equiv (2\pi \ls)^7\,\ls\,\gs^2$ and the field strengths are defined by
\begin{align}
 \hat{G}^{(p)} &\equiv \rmd \hat{C}^{(p-1)} + \hat{H}^{(3)}\wedge \hat{C}^{(p-3)}\quad (1\leq p \leq 5) \,,
\end{align}
and we choose the asymptotic value of the dilation $\hat{\phi}$ to be zero; 
$\Exp{\hat{\phi}}\equiv \gs^{-1}\Exp{\hat{\Phi}}$ ($\hat{\Phi}$: ten-dimensional dilaton).
The Bianchi identities are given by
\begin{align}
 \rmd \hat{H}^{(3)} =0\,,\quad \rmd \hat{G}^{(p)}=0\quad (p=1,2)\,,\quad 
 \rmd \hat{G}^{(p)} + \hat{H}^{(3)}\wedge \hat{G}^{(p-2)} = 0 \quad (p=3,\,4)\,,
\end{align}
and the equations of motion become
\begin{align}
 \text{\underline{IIA:}}\quad &\rmd \bigl(\Exp{-2\hat{\phi}}*_{10}\hat{H}^{(3)}\bigr) + \hat{G}^{(2)}\wedge *_{10} \hat{G}^{(4)}
  + \frac{1}{2}\, \hat{G}^{(4)}\wedge \hat{G}^{(4)} =0 \,,
\\
 &\rmd \hat{G}^{(p)} + \hat{H}^{(3)}\wedge \hat{G}^{(p-2)}=0 \quad (p=6,\,8) \,,
\\
 \text{\underline{IIB:}}\quad &\rmd \bigl(\Exp{-2\hat{\phi}}*_{10}\hat{H}^{(3)}\bigr) + \hat{G}^{(1)}\wedge *_{10}\hat{G}^{(3)}
  - \hat{G}^{(5)}\wedge \hat{G}^{(3)} =0\,,
\\
 &\rmd \hat{G}^{(p)} + \hat{H}^{(3)}\wedge \hat{G}^{(p-2)}=0 \quad (p=5,\,7,\,9) \,,
\end{align}
where $\hat{G}^{(p)}$ for $p=6,\dotsc,9$ is defined 
by the following relation (which is valid for any $p$)
\begin{align}
 \hat{G}^{(p)} = (-1)^{\frac{p(p-1)}{2}}\,*_{10}\hat{G}^{(10-p)}\,.
\end{align}
The equation of motion for the $B$-field can be written 
as the Bianchi identity associated with $\hat{B}^{(6)}$ (i.e., $\rmd^2\hat{B}^{(6)}=0$), 
if we define the dual field $\hat{B}^{(6)}$ as
\begin{align}
 \text{\underline{IIA:}}\quad & 
 \rmd \hat{B}^{(6)}\equiv \Exp{-2\hat{\phi}}*_{10}\hat{H}^{(3)} -\frac{1}{2}\,\hat{G}^{(2)}\wedge \hat{C}^{(5)} 
                          +\frac{1}{2}\,\hat{G}^{(4)}\wedge \hat{C}^{(3)} +\frac{1}{2}\,*_{10} \hat{G}^{(4)}\wedge \hat{C}^{(1)} \,,
\\
 \text{\underline{IIB:}}\quad & 
 \rmd \hat{B}^{(6)}\equiv \Exp{-2\hat{\phi}}*_{10}\hat{H}^{(3)} -\hat{C}^{(4)}\wedge\rmd\hat{C}^{(2)}
                          +\frac{1}{2}\,\hat{C}^{(2)}\wedge \hat{C}^{(2)}\wedge \hat{H}^{(3)} + \hat{C}^{(0)}\, *_{10} \hat{G}^{(3)} \,.
\end{align}

\paragraph{\underline{Duality rules}\\}

If we write the background fields as
\begin{align}
 \rmd s^2&=\widetilde{G}_{\tilde{M}\tilde{N}}\,\rmd x^{\tilde{M}}\,\rmd x^{\tilde{N}}
          +\hat{G}_{yy}\,\bigl(\rmd y+\hat{A}\bigr)^2\,,
\\
 \hat{B}^{(2)}&= \hat{B}^{(2,2)} 
 + \hat{B}^{(2,1)}\wedge \bigl(\rmd y+\hat{A}\bigr) \,,
\\
 \hat{C}^{(p)}&= \hat{C}^{(p,p)} 
 + \hat{C}^{(p,p-1)}\wedge \bigl(\rmd y+\hat{A}\bigr) \,,
\end{align}
the fields after we perform T-duality along the $y$-direction are given by
\begin{align}
 \widetilde{G}'_{\tilde{M}\tilde{N}} &= \widetilde{G}_{\tilde{M}\tilde{N}} \,,\quad
 \hat{G}'_{yy} = \frac{1}{\hat{G}_{yy}} \,,\quad
 \hat{A}' = \hat{B}^{(2,1)}\,,
\\
 \Exp{2\hat{\phi}'}&= \frac{\Exp{2\phi}}{\hat{G}_{yy}} \,,
\quad
 \hat{B}^{\prime (2)} = \hat{B}^{(2,2)} + \hat{A}\wedge \rmd y \,,
\\
 \hat{C}^{\prime (p)}&= \hat{C}^{(p+1,p)} + \hat{C}^{(p-1,p-1)}\wedge(\rmd y+\hat{B}^{(2,1)}) \,.
\end{align}
In addition, the radius and fundamental constants transform as
\begin{align}
 R'_y &= \frac{\ls^2}{R_y} \,, \quad 
 \gs' = \gs\, (\ls/R_y) \,,\quad
 \ls' = \ls \,. 
\label{eq:T-dual-units}
\end{align}

The $SL(2)$-duality transformation rule in type IIB theory is given by
\begin{align}
 \tau' &= \frac{\SLa\,\tau+\SLb}{\SLc\,\tau+\SLd} \qquad 
 \bigl(\tau \equiv \hat{\cC}^{(0)} + \ii \Exp{-\hat{\Phi}} \bigr) \,, 
\\
 \bpm \hat{\cC}^{\prime (2)} \cr \hat{B}^{\prime (2)}\epm
 &=\bpm \SLa & -\SLb \cr -\SLc & \SLd \epm
   \bpm \hat{\cC}^{(2)} \cr \hat{B}^{(2)}\epm \,, \\
 \hat{\cC}^{\prime (4)} 
 &= \hat{\cC}^{(4)} + \hat{B}^{(2)}\wedge \hat{\cC}^{(2)}\,,\quad 
 \hat{G}_{MN}' = \abs{\SLc\,\tau+\SLd}\, \hat{G}_{MN}\,, 
\\
 *'_{10}(\text{$p$-form}) &= \abs{\SLc\,\tau+\SLd}^{5-p}\,*_{10}(\text{$p$-form})\,,
\end{align}
where $\cC^{(p)}\equiv \gs^{-1}\,C^{(p)}$ and $\SLa\SLd-\SLb\SLc=1$. 

In particular, the S-duality transformation rule is given by 
the $SL(2)$ transformation with $\SLa=0=\SLd$, $\SLc=-\SLb=1$, 
followed by a rescaling of background fields 
$\hat{G}_{MN}\to \gs\,\hat{G}_{MN}$\,,\ 
$\hat{B}^{(2)}\to \gs\,\hat{B}^{(2)}$\,, 
$\hat{\cC}^{(2)}\to \gs\,\hat{\cC}^{(2)}$\,, and 
$\hat{\cC}^{(4)}\to \gs^2\,\hat{\cC}^{(4)}$, 
where $\gs$ is the string coupling constant before the $SL(2)$ transformation. 
In total, the transformation rule for the background fields is given by
\begin{align}
  \boldsymbol{\tau}' = -1/\boldsymbol{\tau}\,,\quad 
  \hat{G}_{\mu\nu}' = \abs{\boldsymbol{\tau}}\, \hat{G}_{\mu\nu}\,, \quad 
  \hat{B}^{\prime (2)} = - \hat{C}^{(2)} \,,\quad 
  \hat{C}^{\prime (2)}= \hat{B}^{(2)} \,,
\end{align}
where $\boldsymbol{\tau}\equiv \gs\,\tau = C^{(0)}+\ii\Exp{-\ii\hat{\phi}}$\,.
For the fundamental constants, it is given by
\begin{align}
 \gs' = \frac{1}{\gs} \,, \quad 
 \ls' =\gs^{1/2}\, \ls \,.
\end{align}
Note that the latter follows from the requirement that 
the action for type IIB supergravity is invariant under the above rescaling of the fields.

\section{Electric-magnetic duality in four dimension}
\label{app:ele-mag}

We here consider the NS-NS sector of supergravity in ten dimensions:
\begin{align}
 \frac{1}{2\kappa_{10}^2}\,\int \rmd^{10}x\,\sqrt{-\hat{G}}\Exp{-2\hat{\phi}}
 \Bigl(\hat{R}+4\,\hat{G}^{MN}\,\partial_{M}\hat{\phi}\,\partial_{N}\hat{\phi} 
       -\frac{1}{12}\,\hat{H}^{(3)}_{MNL}\,\hat{H}^{(3)MNL} \Bigr) \,.
\end{align}
We then consider a compactification on a six-torus $T^6$, 
and decompose the metric $\hat{G}_{MN}$ as
\begin{align}
 \bigl(\hat{G}_{MN}\bigr) =
 \bpm
 g_{\mu\nu}+A_{\mu}^{\alpha}\,G_{\alpha\beta}\,A^{\beta}_{\nu}~ 
 & A^{\gamma}_{\mu}\,G_{\gamma\beta}\cr
 G_{\alpha\gamma}\,A^{\gamma}_{\nu}& G_{\alpha\beta}
 \epm
 \,,\quad 
 \bigl(\hat{G}^{MN}\bigr) =
 \bpm
 g^{\mu\nu} & -g^{\mu\rho}\,A_\rho^{\beta} \cr
 -A_\rho^{\alpha}\,g^{\rho\nu}& ~G^{\alpha\beta}+A^{\alpha}_{\mu}\,g^{\mu\nu}\,A_\nu^{\beta}
 \epm \,.
\end{align}
Here, indices $\alpha,\beta,\cdots$ are for coordinates of the internal manifold, 
while indices $\mu,\nu,\cdots$ are for four-dimensional coordinates. 

The four-dimensional action becomes
\begin{align}
 \frac{1}{2\kappa_{4}^2}\,\int \rmd^4 x\,\sqrt{-g}\,\Exp{-2\phi} 
  \Bigl[&R+4\,g^{\mu\nu}\,\partial_{\mu}\phi\,\partial_{\nu}\phi 
  +\frac{1}{4}\,g^{\mu\nu}\,\partial_{\mu}G_{\alpha\beta}\,\partial_{\nu}G^{\alpha\beta} 
  -\frac{1}{4}\,G_{\alpha\beta}\,F_{\mu\nu}^{\alpha}\, F^{\beta\,\mu\nu} 
\nn\\
 &-\frac{1}{4}\,H_{\mu\alpha\beta}\,H^{\mu\alpha\beta}
 -\frac{1}{4}\,H_{\mu\nu\alpha}\,H^{\mu\nu\alpha}
 -\frac{1}{12}\,H_{\mu\nu\rho}\,H^{\mu\nu\rho}\Bigr] \,,
\end{align}
if we define $2\kappa_{4}^2 \equiv 2\kappa_{10}^2/V_{4\cdots 9}$ and
\begin{align}
 F_{\mu\nu}^{\alpha} &\equiv \partial_{\mu}A^{\alpha}_{\nu}-\partial_{\nu}A^{\alpha}_{\mu}\,,\qquad 
 \Exp{2\phi} \equiv \Exp{2\hat{\phi}}/ (\det G_{\alpha\beta})^{1/2}
 \,,
\\
 H_{\mu\alpha\beta}&\equiv\hat{H}^{(3)}_{\mu\alpha\beta} \,,\qquad 
 H_{\mu\nu\alpha}\equiv \hat{H}^{(3)}_{\mu\nu\alpha}
 -2\,A_{[\mu}^{\beta}\hat{H}^{(3)}_{\nu]\alpha\beta}\,,
\\
 H_{\mu\nu\rho}
 &\equiv \hat{H}^{(3)}_{\mu\nu\rho} -3\,A_{[\mu}^{\alpha}\,\hat{H}^{(3)}_{\nu\rho]\alpha}
         +3\,A_{[\mu}^{\alpha}\,A_{\nu}^{\beta}\,\hat{H}^{(3)}_{\rho]\alpha\beta} \,.
\end{align}
In addition, we define
\begin{align}
 B_{\alpha\beta} &\equiv \hat{B}^{(2)}_{\alpha\beta} \,,
\qquad
 A_{\alpha\mu} \equiv 
 \hat{B}^{(2)}_{\mu\alpha}+\hat{B}^{(2)}_{\alpha\beta}\,A_{\mu}^{\beta} \,,\qquad 
 F_{\alpha\mu\nu} \equiv \partial_\mu A_{\alpha\nu}-\partial_\nu A_{\alpha\mu}\,, 
\\
 B_{\mu\nu}
 &\equiv \hat{B}^{(2)}_{\mu\nu} 
 + \frac{1}{2}\,\bigl(A^{\alpha}_{\mu}\,A_{\alpha\nu} 
 - A^{\alpha}_{\nu}\,A_{\alpha\mu} \bigr)
 - \hat{B}^{(2)}_{\alpha\beta}\,A^{\alpha}_{[\mu}\,A_{\nu]}^{\beta} \,,
\end{align}
and then by using $\hat{H}^{(3)} =\rmd \hat{B}^{(2)}$ we obtain
\begin{align}
 H_{\mu\alpha\beta}&=\partial_{\mu}B_{\alpha\beta} \,, 
\qquad 
 H_{\mu\nu\alpha} =F_{\alpha\mu\nu} -B_{\alpha\beta}F^{\beta}_{\mu\nu} \,,
\\
 H_{\mu\nu\rho}
 &= 3\,\partial_{[\mu}B_{\nu\rho]}-\frac{3}{2}\,\LL_{IJ}\,\cA^I_{[\mu}\,\cF^J_{\nu\rho]} 
\qquad (\cF^I\equiv \rmd \cA^I)
\,.
\end{align}
Here, the gauge fields are collected into 
$\bigl(\cA^I_{\mu}\bigr)\equiv (A^{\alpha}_{\mu},A_{\alpha\mu})$ $(I=1,2,\ldots,12)$\,,
and $\LL_{IJ}$ is the $O(6,6)$-invariant metric defined by
\begin{align}
 (\LL_{IJ})\equiv \bpm \mathbf{0}& \mathbf{1} \cr \mathbf{1} & \mathbf{0} \epm 
 \equiv (\LL^{IJ}) \,.
\end{align}
In addition, if we define a matrix
\begin{align}
 M &=\bpm G^{-1}& -G^{-1}\,B \cr B\,G^{-1} & G-B\,G^{-1}\,B \epm 
 \in {\rm SO}(6,6)\,, 
\\
 M^{-1} &=\bpm G-B\,G^{-1}\,B & B\,G^{-1} \cr -G^{-1}\,B & G^{-1} \epm =\LL\,M\,\LL \,,
\end{align}
the four-dimensional action becomes
\begin{align}
 \frac{1}{2\kappa_{4}^2}\,\int \rmd^4x\,\sqrt{-g}\,\Exp{-2\phi}
 \Bigl[&R+4\,g^{\mu\nu}\,\partial_{\mu}\phi\,\partial_{\nu}\phi
        -\frac{1}{12}\,H_{\mu\nu\rho}\,H^{\mu\nu\rho} 
\nn\\
 &-\frac{1}{4}(M^{-1})_{IJ}\,\cF^I_{\mu\nu}\,\cF^{J\mu\nu}
 +\frac{1}{8}\,\Tr \bigl(g^{\mu\nu}\,\partial_{\mu} M\,\partial_{\nu} M^{-1}\bigr)\Bigr]\,.
\end{align}
In the four-dimensional Einstein frame 
$g_{\mu\nu}^{\rm E}\equiv \Exp{-2\phi}g_{\mu\nu}$, 
the action takes the following form:
\begin{align}
 \frac{1}{2\kappa_{4}^2}\,
 \int \rmd^4x\,\sqrt{-g_{\rm E}}\,
 \Bigl[&R_{\rm E}-2\,g^{\mu\nu}\,\partial_\mu \phi\,\partial_\nu \phi 
 -\frac{\Exp{-4\phi}}{12}\,H_{\mu\nu\rho}\,H^{\mu\nu\rho} 
\nn\\
 &-\frac{\Exp{-2\phi}}{4}(M^{-1})_{IJ}\,\cF^I_{\mu\nu}\,\cF^{J\mu\nu}
 +\frac{1}{8}\,\Tr \bigl(g^{\mu\nu}\,\partial_{\mu} M\,\partial_{\nu} M^{-1}\bigr)\Bigr] \,.
\end{align}
The equations of motion of the theory are equivalent 
to those obtained by the action,
\begin{align}
 \frac{1}{2\kappa_{4}^2}\,
 \int \rmd^4x\,\sqrt{-g_{\rm E}}\,
 \Bigl[&R_{\rm E}-2\,g^{\mu\nu}\,\partial_{\mu}\phi\,\partial_{\nu}\phi 
       -\frac{\Exp{4\phi}}{2}\,g^{\mu\nu}\,\partial_{\mu}\axion\,\partial_{\nu}\axion 
    +\frac{1}{8}\,\Tr\bigl(g^{\mu\nu}\,\partial_{\mu} M\,\partial_{\nu} M^{-1}\bigr)
\nn\\
 &-\frac{\Exp{-2\phi}}{4}\,(M^{-1})_{IJ}\,\cF^I_{\mu\nu}\,\cF^{J,\mu\nu} 
  -\frac{1}{4}\,\axion\, \LL_{IJ}\,\cF^{I}_{\mu\nu}\,\tilde{\cF}^{J,\mu\nu}
 \Bigr] ,
\end{align}
where we defined
\begin{align}
 H^{(3)} = - \Exp{4\phi}*_{\rm 4E}\,\rmd \axion \,,\qquad 
 \tilde{\cF}^{(2)} \equiv *_{\rm 4E} \cF^{(2)} \,.
\label{eq:H3-axion}
\end{align}
Finally, defining the axion-dilaton,
\begin{align}
 \lambda \equiv \lambda_1 + \ii\lambda_2 \equiv \axion+\ii \Exp{-2\phi}\,,
\end{align}
we obtain the following four-dimensional action:
\begin{align}
 S_{\rm 4d}=\frac{1}{2\kappa_{4}^2}\,
 \int \rmd^4x\sqrt{-g_{\rm E}}\,
 \Bigl[&R_{\rm E}
 -\frac{g^{\mu\nu}\,\partial_{\mu}\lambda\,\partial_{\nu}\bar{\lambda}}{2\lambda_2^2} 
 +\frac{1}{8}\,\Tr \bigl(g^{\mu\nu}\,\partial_{\mu} M\,\partial_{\nu} M^{-1}\bigr)
\nn\\
 &-\frac{1}{4}\,\lambda_2\,(M^{-1})_{IJ}\,\cF^I_{\mu\nu}\,\cF^{J,\mu\nu} 
  -\frac{1}{4}\,\lambda_1\,\LL_{IJ}\,\cF^I_{\mu\nu}\,\tilde{\cF}^{J,\mu\nu}
 \Bigr] \,.
\label{eq:4d-action}
\end{align}

If we define
\begin{align}
 \cG_I 
 \equiv - \Exp{-2\phi}\,(M^{-1})_{IJ}\, *_{\rm 4E} \cF^{J} 
        - \axion\,\LL_{IJ} \, \cF^{J}\,,
\label{Gpm=NFpm}
\end{align}
the equations of motion and the Bianchi identities (with source terms) become
\begin{align}
 \rmd \bpm  \cF^I \cr \cG_I\epm 
  = -2\kappa_{4}^2\, \sum_p \delta^4 (x-x_p) \,\bpm p^I_p\cr q_{I}^p\epm \,,\quad 
 \bpm p^I \cr q_{I} \epm
  \equiv  \bpm -\frac{1}{2\kappa_{4}^2}\,\int_{S^2} \cF^I \cr -\frac{1}{2\kappa_{4}^2}\,\int_{S^2} \cG_I \epm \,.
\label{eq:4dcharge-def}
\end{align}
This set of equations is invariant under 
the $SL(2,\lZ)$ transformation
\begin{align}
 \bpm \cF^{\prime I} \cr \cG'_I\epm 
  &= \bpm 
  \SLd\,\delta^I{}_J& -\SLc\, \LL^{IJ}\cr 
  -\SLb\,\LL_{IJ}&\SLa\,\delta_I{}^J\cr \epm \, 
  \bpm \cF^J \cr \cG_J\epm \,,
\label{eq:F-G-SL2}
\\
 \lambda'&=\frac{\SLa\,\lambda +\SLb}{\SLc\,\lambda + \SLd}\qquad
 \bigl(\SLa\,\SLd-\SLb\,\SLc=1\bigr)\,,
\end{align}
which is called the electric-magnetic duality of the theory. 
The duality transformation rule for $\cF^I$ can be written as
\begin{align}
 \cF^{\prime I} = (\SLc\,\lambda_1+\SLd)\,\cF^I + \SLc\,\lambda_2\,\LL^{IJ}\,(M^{-1})_{JK}\,*_{\rm 4E}\cF^K \,.
\end{align}

\section{Page charges}
\label{app:Page}

Page charge \cite{Page:1984qv} (see also \cite{Marolf:2000cb,deBoer:2012ma}) 
is one of the possible charge definitions for theories with Chern-Simon terms, 
such as supergravity that describes string theory at low energy. 
We here obtain the expression of Page charges for various defect branes 
which are extending or smeared along a seven-torus $T^7_{3456789}$\,. 
As is the case of the D7-brane, the flux integral for defect branes can always be written as 
an integral of a 1-form along a closed curve which encloses the vortex once. 

The Page current for D$p$-brane is given by (see e.g., \cite{deBoer:2012ma})
\begin{align}
 (2\pi\ls)^{7-p}\,\gs\,*_{10}j_{{\DD}p}
 = \rmd\bigl[\Exp{\hat{B}^{(2)}}\mathbb{G}\bigr]^{(8-p)}
 = \rmd\bigl(\hat{G}^{(8-p)} + \hat{B}^{(2)}\wedge \hat{G}^{(6-p)} + \cdots\bigr) \,,
\label{eq:Page-charge-Dp}
\end{align}
where we introduced the polyform $\mathbb{G}\equiv \hat{G}^{(1)}+\cdots +\hat{G}^{(9)}$ 
and the superscript $(8-p)$ on the square bracket represents extracting the $(8-p)$-form part.%%%
\footnote{In the presence of NS5-brane sources, we should include additional terms to $\mathbb{G}$ which have support only on the NS5-brane worldvolume \cite{deBoer:2012ma}, although we are here assuming the absence of such terms.} 
Since Page charges obey the Dirac quantization condition, 
it will be natural to assume that Page currents change covariantly 
under the action of $T$- and $S$-dualities. 
Namely, for example, under the action of the $T_3$-duality, 
the Page current for D7(3456789)-brane, $*_{10}j_{\DD7}$, 
will become that for (smeared) D6(456789)-brane, $*_{10}j_{\DD6}$. 
In the following, starting from the Page current for D7(3456789)-brane, 
we obtain Page currents for various branes by taking dualities. 

The Page current for D7(3456789)-brane 
and the Page charge contained in a region $V\in \lR^2$ 
($\lR^2$: $x^1$-$x^2$ plane or $z$-plane) are given by
\begin{align}
 *_{10}j_{\DD7}
 &= \gs^{-1}\, \rmd \hat{G}^{(1)}
  = \sigma_{\DD7}^{-1}\, \rmd^2 \hat{C}^{(0)}
 \qquad (\sigma_{\DD7}\equiv \gs) \,,
\\
 Q_{\DD7} &= \int_V *_{10}j_{\DD7} 
 = \sigma_{\DD7}^{-1}\,\int_{\partial V} \rmd \hat{C}^{(0)} \,.
\end{align}
The value of the dimensionless quantity $\sigma_{\DD7}$ changes 
according to the duality transformation rules, 
and it takes the following form for various defect branes which appear in this paper:
\begin{align}
 \sigma_{{\DD}p(i_1\cdots i_p)}
 &= \gs\,(2\pi\ls)^{7-p}\,\frac{V_{i_1\cdots i_p}}{V_{3\cdots 9}}\,,\quad 
 \sigma_{\NS7}= \gs^{-1}\,,\quad 
 \sigma_{6^1_3(456789,3)}= \frac{2\pi R_3}{\gs\,(2\pi\ls)}\,,
\nn\\
 \sigma_{\NS5(56789)}&=\frac{(2\pi\ls)^2}{V_{34}}\,,\quad
 \sigma_{\KKM(56789,i)}=\frac{(2\pi R_i)^2}{V_{34}}\quad (i=3\text{ or }4)\,, 
\nn\\
 \sigma_{5^2_2(56789,34)}&=\frac{V_{34}}{(2\pi\ls)^2}\,, \quad 
 \sigma_{5^2_3(56789,34)}=\frac{V_{34}}{\gs\,(2\pi\ls)^2}\,,\quad 
 \sigma_{\FF1(i)}= \frac{\gs^2\,(2\pi\ls)^6\,(2\pi R_i)}{V_{3\cdots 9}}\,.
\label{eq:various-sigma}
\end{align}
As long as no confusion arises, 
we will use the simple notation such as $\sigma_{\KKM}$ or $\sigma$\,. 
Note that $\sigma_{\KKM(56789,4)}$, $\sigma_{5^2_2(56789,34)}$, and $\sigma_{\DD7}$ 
are the same with $\sigma$'s defined in the bulk of this paper. 

Now, by taking $T_3$-duality, we obtain
\begin{align}
 *_{10}j_{\DD6} = \sigma_{\DD6}^{-1} \, \rmd^2 \iota_3 \hat{C}^{(1)} 
 = \sigma_{\DD6}^{-1}\, \iota_3 \rmd \hat{G}^{(2)}\,,
\end{align}
where $\iota_i$ represents the interior product of the coordinate basis 
$\partial_i$ ($i=3,\dotsc,9$) with differential forms, 
and we used the fact that 
$\iota_i$ anti-commutes with $\rmd$ since $\partial_i$ is a Killing vector. 
The Page charge is thus given by
\begin{align}
  Q_{\DD6} = \int_V *_{10}j_{\DD6} 
 = \sigma_{\DD6}^{-1}\,\int_{\partial V} \rmd \iota_3\hat{C}^{(1)} \,.
\end{align}
By further taking $T_4$-duality, we obtain
\begin{align}
 *_{10}j_{\DD5}
 &= \sigma_{\DD5}^{-1} \, 
     \rmd^2 \iota_3 \bigl[(-\iota_4) \hat{C}^{(2)} + \hat{C}^{(0)}\,(\rmd x^4 -\iota_4 \hat{B}^{(2)}) \bigr] 
\nn\\
 &= \sigma_{\DD5}^{-1} \, 
    \rmd^2  \iota_4\,\iota_3\bigl[\hat{C}^{(2)} + \hat{C}^{(0)}\,\hat{B}^{(2)} \bigr]
 = \sigma_{\DD5}^{-1} \, \iota_4\iota_3\rmd \bigl(\hat{G}^{(3)}+\hat{B}^{(2)}\wedge \hat{G}^{(1)}\bigr)\,,
\end{align}
where we used $\iota_3 \rmd x^4=0$ and $\{\iota_i,\iota_j\}=0$. 
By repeating this kind of calculations, 
the Page current and Page charge for the defect D$p((10-p)\cdots 9)$-brane become
\begin{align}
 *_{10}j_{{\DD}p}
 &= \sigma_{{\DD}p}^{-1}\, \rmd^2
    \bigl[\iota_{9-p}\cdots \iota_3\Exp{\hat{B}^{(2)}}\mathbb{C} \bigr]^{(0)} 
\nn\\
 &= \sigma_{{\DD}p}^{-1}\, 
   \iota_{9-p}\cdots \iota_3 \rmd\bigl[\Exp{\hat{B}^{(2)}}\mathbb{G} \bigr]^{(8-p)} \,,
\\
 Q_{{\DD}p} &= \int_V *_{10}j_{{\DD}p} 
 = \sigma_{{\DD}p}^{-1}\, \int_{C} \rmd\bigl[\iota_{9-p}\cdots \iota_3\Exp{\hat{B}^{(2)}}\mathbb{C} \bigr]^{(0)} 
\nn\\
 &= \frac{1}{(2\pi\ls)^{7-p}\,\gs}\,
   \int_{C\times T_{3\cdots (9-p)}}
   \bigl[\Exp{\hat{B}^{(2)}}\mathbb{G} \bigr]^{(8-p)}\,.
\end{align}
In the last expression, we replaced the interior product by the integral on the $(7-p)$-torus 
(divided by its volume) over which the D$p$-brane is smeared out.
This expression is similar to the known result 
for the (unsmeared) D$p$-brane \eqref{eq:Page-charge-Dp}. 

For the D$p$-brane, the continuity equation for the Page current, 
or the Bianchi identity, can be written as 
[recall that the D$p$-brane is smeared in $(7-p)$ directions]
\begin{align}
  \rmd *_{10}j_{{\DD}p} = \sum_i \delta^2(x-x_i)\,\rmd x^1\wedge\rmd x^2 \,,
\end{align}
where we have included the source terms on the right hand side, 
which violates the Bianchi identity. 
Since the right hand side is invariant under the duality transformations, 
this kind of continuity equation holds for any Page currents derived below. 

Now, considering the $S$-dual of $*_{10}j_{\DD7}$, $*_{10}j_{\DD5}$, and $*_{10}j_{\DD1}$, 
we obtain the following Page currents for NS7-brane, NS5(56789)-brane in type IIB theory, 
and F1(9)-brane in type IIB theory:
\begin{align}
 *_{10}j_{\NS7}
 &= -\sigma_{\NS7}^{-1}\,
   \rmd^2 \biggl(\frac{\hat{C}^{(0)}}{\abs{\boldsymbol{\tau}}^2}\biggr)\,,
\quad
 *_{10}j_{\NS5}
  = \sigma_{{\NS}5}^{-1} \rmd^2 \biggl(\hat{B}^{(2)}_{34} + \hat{C}^{(2)}_{34}\,\frac{\hat{C}^{(0)}}{\abs{{\boldsymbol\tau}}^2}\biggr)\,,
\label{eq:Page-NS5}
\\
 *_{10}j_{\FF1}
 &=  \sigma_{\FF1}^{-1}\, \iota_8\cdots \iota_3 \,
 \rmd\biggl[\Exp{-2\hat{\phi}}*_{10}\hat{H}^{(3)} + C^{(0)}\,*_{10}\hat{G}^{(3)}
       +C^{(2)}\wedge \hat{G}^{(5)}
\nn\\
 &\qquad\qquad\qquad\qquad  - \frac{1}{2}\,\hat{C}^{(2)} \wedge \hat{C}^{(2)} \wedge \hat{H}^{(3)}
 - \rmd \Bigl(\frac{\hat{C}^{(0)}\,\hat{C}^{(2)} \wedge\hat{C}^{(2)} \wedge\hat{C}^{(2)}}{6\,\abs{\boldsymbol{\tau}}^2}\Bigr)
 \biggr] 
\nn\\
 &= \sigma_{\FF1}^{-1}\, 
 \rmd^2\iota_8\cdots \iota_3 \biggl[ \hat{B}^{(6)} +  \hat{C}^{(4)} \wedge \hat{C}^{(2)} 
 - \frac{\hat{C}^{(0)}\,\hat{C}^{(2)} \wedge\hat{C}^{(2)} \wedge\hat{C}^{(2)}}{6\,\abs{\boldsymbol{\tau}}^2} \biggr] \,,
\label{eq:PageF1}
\end{align}
where $\boldsymbol{\tau}=\gs\,\tau= C^{(0)} + \ii \Exp{-\hat\phi}$.
Note that the expression for $*_{10}j_{\NS5}$ is not invariant under 
the action of $T$-dualities in the $x^5,\dotsc,x^9$-directions 
while the NS5(56789)-brane is invariant under these $T$-dualities. 
Although we do not have a full understanding on this point, 
the difference in the expression does not change the actual value 
of the charge for typical examples. 

By taking $T_4$-duality for $*_{10}j_{\NS5}$, 
we obtain the Page current for KKM(56789,4)-brane in type IIA theory;
\begin{align}
 *_{10}j_{\KKM}
 &= -\sigma_{{\KKM}}^{-1} \, \rmd^2 
 \biggl( \frac{\hat{G}_{34}+\Exp{2\hat{\phi}}\,\hat{C}^{(1)}_3\,\hat{C}^{(1)}_4}{\hat{G}_{44}+\Exp{2\hat{\phi}}\,\bigl(\hat{C}_4^{(1)}\bigr)^2} \biggr) \,. 
\end{align}
By taking $T_3$-duality further, we obtain the following expression for $5^2_2(56789,34)$-brane in type IIB theory:
\begin{align}
 *_{10}j_{5^2_2}
 &= \sigma_{5^2_2}^{-1}\, \rmd^2 
 \biggl(\frac{\hat{B}^{(2)}_{34}+\Exp{2\hat{\phi}}\,\hat{C}^{(0)}\,\bigl(\hat{C}^{(2)}_{34}+\hat{B}^{(2)}_{34}\,\hat{C}^{(0)}\bigr)}
             {\det\bigl(\hat{G}_{ab}+\hat{B}^{(2)}_{ab}\bigr) +\Exp{2\hat{\phi}}\,\bigl(\hat{C}^{(2)}_{34}+\hat{B}^{(2)}_{34}\,\hat{C}^{(0)}\bigr)^2} \biggr) \,,
\end{align}
where $\bigl(\hat{G}_{ab}+\hat{B}^{(2)}_{ab}\bigr)
\equiv \Bigl(\begin{smallmatrix}\hat{G}_{33} & \hat{G}_{34}+\hat{B}^{(2)}_{34}\cr
\hat{G}_{34}-\hat{B}^{(2)}_{34} & \hat{G}_{44} \end{smallmatrix}\Bigr)$\,. 
In particular, in the absence of the Ramond-Ramond fields, we simply have
\begin{align}
 *_{10}j_{5^2_2}
 &= \sigma_{5^2_2}^{-1}\, \rmd^2 
 \biggl(\frac{\hat{B}^{(2)}_{34}}{\det\bigl(\hat{G}_{ab}+\hat{B}^{(2)}_{ab}\bigr)} \biggr) \,.
\label{eq:Page-522}
\end{align}

Finally, we will comment on the relation between 
the Page current $*_{10}j_{5^2_2}$ and the $Q$-flux known in the literature 
\cite{Shelton:2005cf,Andriot:2011uh,Chatzistavrakidis:2013jqa,Andriot:2014uda}.\footnote{We would like to thank David Andriot and Andr\'e Betz for useful discussions about this appendix.} 
In the formulation of $\beta$-supergravity \cite{Andriot:2013xca,Chatzistavrakidis:2013jqa}, 
we introduce new fields $(\tilde{g}_{MN},\,\beta^{MN},\,\tilde{\phi})$ by
\begin{align}
 &\tilde{g} 
  \equiv \bigl(\hat{G}-\hat{B}\bigr)\,\hat{G}^{-1}\,\bigl(\hat{G}+\hat{B}\bigr) \,,
\quad
 \beta \equiv \bigl(\hat{G}+\hat{B}\bigr)^{-1}\,\hat{B}\,\bigl(\hat{G}-\hat{B}\bigr)^{-1} \,,
\\
 &\cH^{-1} = 
 \bpm 
 \hat{G}^{-1} & - \hat{G}^{-1} \,\hat{B} \cr
 \hat{B} \, \hat{G}^{-1} 
 & \hat{G} -\hat{B} \, \hat{G}^{-1} \,\hat{B} 
 \epm 
 \equiv 
 \bpm 
 \tilde{g}^{-1} -\beta \, \tilde{g} \,\beta & -\beta \,\tilde{g} \cr
  \tilde{g} \, \beta & \tilde{g} 
 \epm  \,,
\\
 &\Exp{-2\tilde{\phi}}\sqrt{\abs{\tilde{g}}}\equiv \Exp{-2\hat{\phi}}\sqrt{\abs{\hat{G}}}\,,
\end{align}
and treat them as fundamental variables. 
This redefinition of background fields enables us 
to describe some non-geometric backgrounds geometrically. 
In this formulation, instead of the usual $H$-flux, 
we can naturally define different fluxes, called $Q$-flux and $R$-flux. 
The field strength $Q^{(1)MN}$, which is associated with the $Q$-flux, 
is defined by
\begin{align}
 Q^{(1)MN} \equiv Q_P^{MN} \rmd x^P 
 \equiv \rmd \beta^{MN} + 2\,\beta^{P[M}\,\partial_P\tilde{e}^{N]}_{~\,A}\,\tilde{e}^A \,,\quad 
 \tilde{e}^A\equiv \tilde{e}^A_{~\,M}\,\rmd x^M\,.
\end{align}
Here, $\tilde{e}^A_{~\,M}$ is the vielbein associated with the ``dual'' metric; 
$\tilde{g}_{MN}=\tilde{e}^A_{~\,M}\,\tilde{e}^B_{~\,N}\,\eta_{AB}$\,, 
and $A, B,\cdots$ are the tangent space indices and $\eta_{AB}$ is the flat metric. 
In the presence of the ``$Q$-brane,'' which is nothing but the $5^2_2$-brane, the field strength satisfies the following Bianchi identity \cite{Andriot:2014uda}:%%%
\footnote{Here, we have rewritten equation (1.12) in \cite{Andriot:2014uda} by using the curved indices $M,N,\cdots$. Note that, in \cite{Andriot:2014uda}, the flat indices are $a,b,\dotsc,l$ while the curved ones are $m,n,\dotsc,z$\,.}
\begin{align}
 &\rmd Q^{(1)MN} 
 + \bigl(2\,Q^{P[M}_C\,\partial_P \tilde{e}^{N]}_{~\,D} 
 - \tilde{e}^M_{~\,A}\,\tilde{e}^N_{~\,B}\,\beta^{P[A}\,\partial_P f^{B]}_{~\,CD} 
 \bigr)\,\tilde{e}^C\wedge \tilde{e}^D 
\nn\\
 &= \sigma_{5^2_2}\,\sum_i \delta^2(x-x_i)\,\rmd x^1\wedge\rmd x^2\quad 
 \bigl(Q^{PM}_C\equiv \tilde{e}^N_{~\,C}\,Q^{PM}_N\bigr)\,,
\end{align}
where the right hand side is the source term 
and the structure constant $f^{A}_{~\,BC}$ is defined by
\begin{align}
 f^{A}_{~\,BC} \equiv -2\,\tilde{e}^M_{~\,[C}\,\partial_{B]}\tilde{e}^A_{~\,M}\,,\quad 
 \bigl[\tilde{e}^Q_{~\,B}\,\partial_Q,\, \tilde{e}^R_{~\,C}\,\partial_R\bigr] 
 = f^{A}_{~\,BC}\,\tilde{e}^P_{~\,A}\,\partial_P \,.
\end{align}
If we assume that $\beta^{MN}$ does not have $M=1,2$ components, 
it follows that $\beta^{MN}\,\partial_N$ acted on any field becomes zero 
since we are now assuming the isometries in $03456789$-directions. 
In this case, we simply have
\begin{align}
 Q^{(1)MN} = \rmd \beta^{MN} \,,\quad 
 \rmd Q^{(1)MN} = \sigma_{5^2_2}\,\sum_i \delta^2(x-x_i)\,\rmd x^1\wedge\rmd x^2\,,
\end{align}
where we used $Q^{MN}_P\,\partial_N=0$, which follows from the above assumption. 

Now, in order to relate the above Bianchi identity with that for the Page charge 
\eqref{eq:Page-522}, 
we assume that there are no mixing between $x^3$-$x^4$ directions 
and the other directions in the generalized metric $\cH^{-1}$\,. 
Then, the $x^3$-$x^4$ components of $\beta$ becomes
\begin{align}
 \bigl(\beta^{ab}\bigr) 
 = \frac{1}{\det\bigl(\hat{G}_{ab}+\hat{B}^{(2)}_{ab}\bigr)}\,
  \bpm 0&\hat{B}^{(2)}_{34}\cr -\hat{B}^{(2)}_{34}&0 \epm \,,
\end{align}
and the above Bianchi identity gives
\begin{align}
 \rmd Q^{(1)34} = \rmd^2 \beta^{34} 
 = \rmd^2\biggl(\frac{\hat{B}^{(2)}_{34}}{\det\bigl(\hat{G}_{ab}+\hat{B}^{(2)}_{ab}\bigr)}\biggr)
 = \sigma_{5^2_2}\,\sum_i \delta^2(x-x_i)\,\rmd x^1\wedge\rmd x^2\,,
\end{align}
which is nothing but the Bianchi identity for the Page charge \eqref{eq:Page-522}.


\newpage

\begin{thebibliography}{99}
%\cite{Hull:2004in}
\bibitem{Hull:2004in}
  C.~M.~Hull,
  ``A geometry for non-geometric string backgrounds,''
  JHEP {\bf 0510}, 065 (2005)
  [arXiv:hep-th/0406102].
  %%CITATION = JHEPA,0510,065;%%

%\cite{deBoer:2010ud}
\bibitem{deBoer:2010ud}
  J.~de Boer and M.~Shigemori,
  ``Exotic branes and non-geometric backgrounds,''
  Phys.\ Rev.\ Lett.\  {\bf 104}, 251603 (2010)
  [arXiv:1004.2521 [hep-th]].
  %%CITATION = PRLTA,104,251603;%%

%\cite{LozanoTellechea:2000mc}
\bibitem{LozanoTellechea:2000mc} 
  E.~Lozano-Tellechea and T.~Ortin,
  ``7-branes and higher Kaluza-Klein branes,''
  Nucl.\ Phys.\ B {\bf 607}, 213 (2001)
  [hep-th/0012051].
  %%CITATION = HEP-TH/0012051;%%

%\cite{deBoer:2012ma}
\bibitem{deBoer:2012ma} 
  J.~de Boer and M.~Shigemori,
  ``Exotic Branes in String Theory,''
  Phys.\ Rept.\  {\bf 532}, 65 (2013)
  [arXiv:1209.6056 [hep-th]].
  %%CITATION = ARXIV:1209.6056;%%

%\cite{Page:1984qv}
\bibitem{Page:1984qv} 
  D.~N.~Page,
  ``Classical Stability of Round and Squashed Seven Spheres in Eleven-dimensional Supergravity,''
  Phys.\ Rev.\ D {\bf 28}, 2976 (1983).
  %%CITATION = PHRVA,D28,2976;%%

%\cite{Marolf:2000cb}
\bibitem{Marolf:2000cb} 
  D.~Marolf,
  ``Chern-Simons terms and the three notions of charge,''
  hep-th/0006117.
  %%CITATION = HEP-TH/0006117;%%

%\cite{Schwarz:1982ec}
\bibitem{Schwarz:1982ec} 
  A.~S.~Schwarz,
  ``Field Theories With No Local Conservation Of The Electric Charge,''
  Nucl.\ Phys.\ B {\bf 208}, 141 (1982).  
%%CITATION = NUPHA,B208,141;%%

%\cite{Alford:1990mk}
\bibitem{Alford:1990mk} 
  M.~G.~Alford, K.~Benson, S.~R.~Coleman, J.~March-Russell and F.~Wilczek,
  ``The Interactions and Excitations of Nonabelian Vortices,''
  Phys.\ Rev.\ Lett.\  {\bf 64}, 1632 (1990)
  [Erratum-ibid.\  {\bf 65}, 668 (1990)].
  %%CITATION = PRLTA,64,1632;%%

%\cite{Preskill:1990bm}
\bibitem{Preskill:1990bm} 
  J.~Preskill and L.~M.~Krauss,
  ``Local Discrete Symmetry and Quantum Mechanical Hair,''
  Nucl.\ Phys.\ B {\bf 341}, 50 (1990).
  %%CITATION = NUPHA,B341,50;%%

%\cite{Blinnikov:1982eh}
\bibitem{Blinnikov:1982eh} 
  S.~I.~Blinnikov and M.~Y.~Khlopov,
  ``On Possible Effects Of 'mirror' Particles,''
  Sov.\ J.\ Nucl.\ Phys.\  {\bf 36}, 472 (1982)
  [Yad.\ Fiz.\  {\bf 36}, 809 (1982)].
  %%CITATION = SJNCA,36,472;%%

%\cite{Brekke:1991ap}
\bibitem{Brekke:1991ap} 
  L.~Brekke, W.~Fischler and T.~D.~Imbo,
  ``Alice strings, magnetic monopoles and charge quantization,''
  Phys.\ Rev.\ Lett.\  {\bf 67}, 3643 (1991).
  %%CITATION = PRLTA,67,3643;%%

%\cite{BenMenahem:1992fe}
\bibitem{BenMenahem:1992fe} 
  S.~Ben-Menahem and A.~R.~Cooper,
  ``Superconductivity solves the monopole problem for Alice Strings,''
  Nucl.\ Phys.\ B {\bf 388}, 393 (1992)
  [hep-ph/9204227].
  %%CITATION = HEP-PH/9204227;%%

%\cite{Kobayashi:2010na}
\bibitem{Kobayashi:2010na} 
  S.~Kobayashi, M.~Kobayashi, Y.~Kawaguchi, M.~Nitta and M.~Ueda,
  ``Topological Influence between Monopoles and Vortices: a Possible Resolution of the Monopole Problem,''
  arXiv:1007.3832 [hep-ph].
  %%CITATION = ARXIV:1007.3832;%%

%\cite{Stephen:1974ur}
\bibitem{Stephen:1974ur} 
  M.~J.~Stephen and J.~P.~Straley,
  ``Physics of liquid crystals,''
  Rev.\ Mod.\ Phys.\  {\bf 46}, 617 (1974).
  %%CITATION = RMPHA,46,617;%%

%\cite{Wright:1989zz}
\bibitem{Wright:1989zz} 
  D.~C.~Wright and N.~D.~Mermin,
  ``Crystalline liquids: the blue phases,''
  Rev.\ Mod.\ Phys.\  {\bf 61}, 385 (1989).
  %%CITATION = RMPHA,61,385;%%

%\cite{Harvey:2007ab}
\bibitem{Harvey:2007ab} 
  J.~A.~Harvey and A.~B.~Royston,
  ``Localized modes at a D-brane-O-plane intersection and heterotic Alice atrings,''
  JHEP {\bf 0804}, 018 (2008)
  [arXiv:0709.1482 [hep-th]].
  %%CITATION = ARXIV:0709.1482;%%

%\cite{Meessen:1998qm}
\bibitem{Meessen:1998qm} 
  P.~Meessen and T.~Ortin,
  ``An Sl(2,Z) multiplet of nine-dimensional type II supergravity theories,''
  Nucl.\ Phys.\ B {\bf 541}, 195 (1999)
  [hep-th/9806120].
  %%CITATION = HEP-TH/9806120;%%

%\cite{Onemli:2000kb}
\bibitem{Onemli:2000kb} 
  V.~K.~Onemli and B.~Tekin,
  ``Kaluza-Klein vortices,''
  JHEP {\bf 0101}, 034 (2001)
  [hep-th/0011287].
  %%CITATION = HEP-TH/0011287;%%

%\cite{Giusto:2010gv}
\bibitem{Giusto:2010gv} 
  S.~Giusto and S.~D.~Mathur,
  ``Unwinding of strings thrown into a fuzzball,''
  JHEP {\bf 1007}, 009 (2010)
  [Erratum-ibid.\  {\bf 1104}, 032 (2011)]
  [arXiv:1004.4142 [hep-th]].
  %%CITATION = ARXIV:1004.4142;%%

%\cite{Hull:2009mi}
\bibitem{Hull:2009mi} 
  C.~Hull and B.~Zwiebach,
  ``Double Field Theory,''  JHEP {\bf 0909}, 099 (2009)  
  [arXiv:0904.4664 [hep-th]]. 
  %%CITATION = ARXIV:0904.4664;%%

%\cite{Hull:2009zb}
\bibitem{Hull:2009zb} 
  C.~Hull and B.~Zwiebach,
  ``The Gauge algebra of double field theory and Courant brackets,'' 
  JHEP {\bf 0909}, 090 (2009)  [arXiv:0908.1792 [hep-th]]. 
  %%CITATION = ARXIV:0908.1792;%%

%\cite{Hohm:2010jy}
\bibitem{Hohm:2010jy} 
  O.~Hohm, C.~Hull and B.~Zwiebach,
  ``Background independent action for double field theory,'' 
  JHEP {\bf 1007}, 016 (2010)  [arXiv:1003.5027 [hep-th]]. 
  %%CITATION = ARXIV:1003.5027;%%

%\cite{Hohm:2010pp}
\bibitem{Hohm:2010pp} 
  O.~Hohm, C.~Hull and B.~Zwiebach,
  ``Generalized metric formulation of double field theory,'' 
  JHEP {\bf 1008}, 008 (2010)  [arXiv:1006.4823 [hep-th]]. 
  %%CITATION = ARXIV:1006.4823;%%

%\cite{Hohm:2011dv}
\bibitem{Hohm:2011dv}
  O.~Hohm, S.~K.~Kwak and B.~Zwiebach,
  ``Double Field Theory of Type II Strings,''
  JHEP {\bf 1109}, 013 (2011)
  [arXiv:1107.0008 [hep-th]].
  %%CITATION = ARXIV:1107.0008;%%

%\cite{Hull:2006va}
  \bibitem{Hull:2006va} 
  C.~M.~Hull,
  ``Doubled Geometry and T-Folds,''
  JHEP {\bf 0707}, 080 (2007)
  [hep-th/0605149].
  %%CITATION = HEP-TH/0605149;%%

%\cite{Gregory:1997te}
\bibitem{Gregory:1997te} 
  R.~Gregory, J.~A.~Harvey and G.~W.~Moore,
  ``Unwinding strings and t duality of Kaluza-Klein and h monopoles,''
  Adv.\ Theor.\ Math.\ Phys.\  {\bf 1}, 283 (1997)
  [hep-th/9708086].
  %%CITATION = HEP-TH/9708086;%%

%\cite{Roy:1998ev}
\bibitem{Roy:1998ev} 
  S.~Roy,
  ``Kaluza-Klein and H dyons in string theory,''
  Phys.\ Rev.\ D {\bf 60}, 082003 (1999)
  [hep-th/9811098].
  %%CITATION = HEP-TH/9811098;%%

%\cite{Schwarz:1993mg}
\bibitem{Schwarz:1993mg} 
  J.~H.~Schwarz and A.~Sen,
  ``Duality symmetries of 4-D heterotic strings,''
  Phys.\ Lett.\ B {\bf 312}, 105 (1993)
  [hep-th/9305185].
  %%CITATION = HEP-TH/9305185;%%

%\cite{Sen:1997zb}
\bibitem{Sen:1997zb} 
  A.~Sen,
  ``Kaluza-Klein dyons in string theory,''
  Phys.\ Rev.\ Lett.\  {\bf 79}, 1619 (1997)
  [hep-th/9705212].
  %%CITATION = HEP-TH/9705212;%%

%\cite{Kimura:2013fda}
\bibitem{Kimura:2013fda} 
  T.~Kimura and S.~Sasaki,
  ``Gauged Linear Sigma Model for Exotic Five-brane,''
  Nucl.\ Phys.\ B {\bf 876}, 493 (2013)
  [arXiv:1304.4061 [hep-th]].
  %%CITATION = ARXIV:1304.4061;%%

%\cite{Striet:2000bf}
\bibitem{Striet:2000bf} 
  J.~Striet and F.~A.~Bais,
  ``Simple models with Alice fluxes,''
  Phys.\ Lett.\ B {\bf 497}, 172 (2000)
  [hep-th/0010236].
  %%CITATION = HEP-TH/0010236;%%

%\cite{Bucher:1992bd}
\bibitem{Bucher:1992bd} 
  M.~Bucher, H.~-K.~Lo and J.~Preskill,
  ``Topological approach to Alice electrodynamics,''
  Nucl.\ Phys.\ B {\bf 386}, 3 (1992)
  [hep-th/9112039].
  %%CITATION = HEP-TH/9112039;%%

%\cite{Kikuchi:2012za}
\bibitem{Kikuchi:2012za} 
  T.~Kikuchi, T.~Okada and Y.~Sakatani,
  ``Rotating string in doubled geometry with generalized isometries,''
  Phys.\ Rev.\ D {\bf 86}, 046001 (2012)
  [arXiv:1205.5549 [hep-th]].
  %%CITATION = ARXIV:1205.5549;%%

%\cite{Kimura:2014wga}
\bibitem{Kimura:2014wga} 
  T.~Kimura,
  ``Defect (p,q) Five-branes,''
  arXiv:1410.8403 [hep-th].
  %%CITATION = ARXIV:1410.8403;%%

%\cite{Kugo:1992md}
\bibitem{Kugo:1992md} 
  T.~Kugo and B.~Zwiebach,
  ``Target space duality as a symmetry of string field theory,''
  Prog.\ Theor.\ Phys.\  {\bf 87}, 801 (1992)
  [hep-th/9201040].
  %%CITATION = HEP-TH/9201040;%%

%\cite{Geissbuhler:2013uka}
\bibitem{Geissbuhler:2013uka} 
  D.~Geissbuhler, D.~Marques, C.~Nunez and V.~Penas,
  ``Exploring Double Field Theory,''
  JHEP {\bf 1306}, 101 (2013)
  [arXiv:1304.1472 [hep-th]].
  %%CITATION = ARXIV:1304.1472;%%

%\cite{Aldazabal:2013sca}
\bibitem{Aldazabal:2013sca} 
  G.~Aldazabal, D.~Marques and C.~Nunez,
  ``Double Field Theory: A Pedagogical Review,''
  Class.\ Quant.\ Grav.\  {\bf 30}, 163001 (2013)
  [arXiv:1305.1907 [hep-th]].
  %%CITATION = ARXIV:1305.1907;%%

%\cite{Hohm:2013bwa}
\bibitem{Hohm:2013bwa} 
  O.~Hohm, D.~L\"ust and B.~Zwiebach,
  ``The Spacetime of Double Field Theory: Review, Remarks, and Outlook,''
  Fortsch.\ Phys.\  {\bf 61}, 926 (2013)
  [arXiv:1309.2977 [hep-th]].
  %%CITATION = ARXIV:1309.2977;%%

%\cite{Hohm:2012gk}
\bibitem{Hohm:2012gk} 
  O.~Hohm and B.~Zwiebach,
  ``Large Gauge Transformations in Double Field Theory,''
  arXiv:1207.4198 [hep-th].
  %%CITATION = ARXIV:1207.4198;%%

%\cite{Fukuma:1999jt}
\bibitem{Fukuma:1999jt} 
  M.~Fukuma, T.~Oota and H.~Tanaka,
  ``Comments on T dualities of Ramond-Ramond potentials on tori,''
  Prog.\ Theor.\ Phys.\  {\bf 103}, 425 (2000)
  [hep-th/9907132].
  %%CITATION = HEP-TH/9907132;%%

%\cite{Bergshoeff:2001pv}
\bibitem{Bergshoeff:2001pv} 
  E.~Bergshoeff, R.~Kallosh, T.~Ortin, D.~Roest and A.~Van Proeyen,
  ``New formulations of D = 10 supersymmetry and D8 - O8 domain walls,''
  Class.\ Quant.\ Grav.\  {\bf 18}, 3359 (2001)
  [hep-th/0103233].
  %%CITATION = HEP-TH/0103233;%%

%\cite{Kimura:2014upa}
\bibitem{Kimura:2014upa} 
  T.~Kimura, S.~Sasaki and M.~Yata,
  ``World-volume Effective Actions of Exotic Five-branes,''
  JHEP {\bf 1407}, 127 (2014)
  [arXiv:1404.5442 [hep-th]].
  %%CITATION = ARXIV:1404.5442;%%

%\cite{Tseytlin:1990nb}
\bibitem{Tseytlin:1990nb} 
  A.~A.~Tseytlin,
  ``Duality Symmetric Formulation of String World Sheet Dynamics,''
  Phys.\ Lett.\ B {\bf 242}, 163 (1990).
  %%CITATION = PHLTA,B242,163;%%

%\cite{Tseytlin:1990va}
\bibitem{Tseytlin:1990va} 
  A.~A.~Tseytlin,
  ``Duality symmetric closed string theory and interacting chiral scalars,''
  Nucl.\ Phys.\ B {\bf 350}, 395 (1991).
  %%CITATION = NUPHA,B350,395;%%

%\cite{Tseytlin:1990hn}
\bibitem{Tseytlin:1990hn} 
  A.~A.~Tseytlin,
  ``Duality symmetric string theory and the cosmological constant problem,''
  Phys.\ Rev.\ Lett.\  {\bf 66}, 545 (1991).
  %%CITATION = PRLTA,66,545;%%
  
%\cite{Copland:2011wx} 
\bibitem{Copland:2011wx} 
  N.~B.~Copland,
  ``A Double Sigma Model for Double Field Theory,''
  JHEP {\bf 1204}, 044 (2012)
  [arXiv:1111.1828 [hep-th]].
  %%CITATION = ARXIV:1111.1828;%%

%\cite{Hohm:2013jma}
\bibitem{Hohm:2013jma} 
  O.~Hohm and H.~Samtleben,
  ``U-duality covariant gravity,''
  JHEP {\bf 1309}, 080 (2013)
  [arXiv:1307.0509 [hep-th]].
  %%CITATION = ARXIV:1307.0509;%%

%\cite{Blair:2013gqa}
\bibitem{Blair:2013gqa} 
  C.~D.~A.~Blair, E.~Malek and J.~H.~Park,
  ``M-theory and Type IIB from a Duality Manifest Action,''
  JHEP {\bf 1401}, 172 (2014)
  [arXiv:1311.5109 [hep-th]].
  %%CITATION = ARXIV:1311.5109;%%

%\cite{Hohm:2013vpa}
\bibitem{Hohm:2013vpa} 
  O.~Hohm and H.~Samtleben,
  ``Exceptional Field Theory I: $E_{6(6)}$ covariant Form of M-Theory and Type IIB,''
  Phys.\ Rev.\ D {\bf 89}, 066016 (2014)
  [arXiv:1312.0614 [hep-th]].
  %%CITATION = ARXIV:1312.0614;%%

%\cite{Hohm:2013uia}
\bibitem{Hohm:2013uia} 
  O.~Hohm and H.~Samtleben,
  ``Exceptional Field Theory II: E$_{7(7)}$,''
  Phys.\ Rev.\ D {\bf 89}, 066017 (2014)
  [arXiv:1312.4542 [hep-th]].
  %%CITATION = ARXIV:1312.4542;%%

%\cite{Park:2014una}
\bibitem{Park:2014una} 
  J.~H.~Park and Y.~Suh,
  ``U-gravity: SL(N),''
  JHEP {\bf 1406}, 102 (2014)
  [arXiv:1402.5027 [hep-th]].
  %%CITATION = ARXIV:1402.5027;%%

%\cite{Godazgar:2014nqa}
\bibitem{Godazgar:2014nqa} 
  H.~Godazgar, M.~Godazgar, O.~Hohm, H.~Nicolai and H.~Samtleben,
  ``Supersymmetric E$_{7(7)}$ Exceptional Field Theory,''
  JHEP {\bf 1409}, 044 (2014)
  [arXiv:1406.3235 [hep-th]].
  %%CITATION = ARXIV:1406.3235;%%

%\cite{Hohm:2014fxa}
\bibitem{Hohm:2014fxa} 
  O.~Hohm and H.~Samtleben,
  ``Exceptional Field Theory III: E$_{8(8)}$,''
  Phys.\ Rev.\ D {\bf 90}, 066002 (2014)
  [arXiv:1406.3348 [hep-th]].
  %%CITATION = ARXIV:1406.3348;%%

%\cite{Hanany:1996ie}
\bibitem{Hanany:1996ie} 
  A.~Hanany and E.~Witten,
  ``Type IIB superstrings, BPS monopoles, and three-dimensional gauge dynamics,''
  Nucl.\ Phys.\ B {\bf 492}, 152 (1997)
  [hep-th/9611230].
  %%CITATION = HEP-TH/9611230;%%

%\cite{Mathur:2005zp}
\bibitem{Mathur:2005zp} 
  S.~D.~Mathur,
  ``The Fuzzball proposal for black holes: An Elementary review,''
  Fortsch.\ Phys.\  {\bf 53}, 793 (2005)
  [hep-th/0502050].
  %%CITATION = HEP-TH/0502050;%%
  
%\cite{Bena:2007kg}
\bibitem{Bena:2007kg} 
  I.~Bena and N.~P.~Warner,
  ``Black holes, black rings and their microstates,''
  Lect.\ Notes Phys.\  {\bf 755}, 1 (2008)
  [hep-th/0701216].
  %%CITATION = HEP-TH/0701216;%%

%\cite{Skenderis:2008qn}
\bibitem{Skenderis:2008qn} 
  K.~Skenderis and M.~Taylor,
  ``The fuzzball proposal for black holes,''
  Phys.\ Rept.\  {\bf 467}, 117 (2008)
  [arXiv:0804.0552 [hep-th]].
  %%CITATION = ARXIV:0804.0552;%%

%\cite{Balasubramanian:2008da}
\bibitem{Balasubramanian:2008da} 
  V.~Balasubramanian, J.~de Boer, S.~El-Showk and I.~Messamah,
  ``Black Holes as Effective Geometries,''
  Class.\ Quant.\ Grav.\  {\bf 25}, 214004 (2008)
  [arXiv:0811.0263 [hep-th]].
  %%CITATION = ARXIV:0811.0263;%%

%\cite{Shelton:2005cf}
\bibitem{Shelton:2005cf} 
  J.~Shelton, W.~Taylor and B.~Wecht,
  ``Nongeometric flux compactifications,''
  JHEP {\bf 0510}, 085 (2005)
  [hep-th/0508133].
  %%CITATION = HEP-TH/0508133;%%

%\cite{Andriot:2011uh}
\bibitem{Andriot:2011uh} 
  D.~Andriot, M.~Larfors, D.~Lust and P.~Patalong,
  ``A ten-dimensional action for non-geometric fluxes,''
  JHEP {\bf 1109}, 134 (2011)
  [arXiv:1106.4015 [hep-th]].
  %%CITATION = ARXIV:1106.4015;%%

%\cite{Chatzistavrakidis:2013jqa}
\bibitem{Chatzistavrakidis:2013jqa} 
  A.~Chatzistavrakidis, F.~F.~Gautason, G.~Moutsopoulos and M.~Zagermann,
  ``Effective actions of non-geometric fivebranes,''
  Phys.\ Rev.\ D {\bf 89}, 066004 (2014)
  [arXiv:1309.2653 [hep-th]].
  %%CITATION = ARXIV:1309.2653;%%

%\cite{Andriot:2014uda}
\bibitem{Andriot:2014uda} 
  D.~Andriot and A.~Betz,
  ``NS-branes, source corrected Bianchi identities, and more on backgrounds with non-geometric fluxes,''
  JHEP {\bf 1407}, 059 (2014)
  [arXiv:1402.5972 [hep-th]].
  %%CITATION = ARXIV:1402.5972;%%

%\cite{Andriot:2013xca}
\bibitem{Andriot:2013xca} 
  D.~Andriot and A.~Betz,
  ``$\beta$-supergravity: a ten-dimensional theory with non-geometric fluxes, and its geometric framework,''
  JHEP {\bf 1312}, 083 (2013)
  [arXiv:1306.4381 [hep-th]].
  %%CITATION = ARXIV:1306.4381;%%

\end{thebibliography}
\end{document}